%% file: main.tex
\UseRawInputEncoding
\documentclass[sigconf]{acmart}

\acmYear{2021}\copyrightyear{2021}
\setcopyright{acmcopyright}
\acmConference[ICS '21]{2021 International Conference on Supercomputing}{June 14--17, 2021}{Virtual Event, USA}
\acmBooktitle{2021 International Conference on Supercomputing (ICS '21), June 14--17, 2021, Virtual Event, USA}
\acmPrice{15.00}
\acmDOI{10.1145/3447818.3460353}
\acmISBN{978-1-4503-8335-6/21/06}

\usepackage{booktabs} 

\usepackage{amsmath,amssymb,amsfonts}
\usepackage{graphicx}
\usepackage{textcomp}
\usepackage{latexsym}
\usepackage[ruled,vlined]{algorithm2e}
\usepackage{tabularx}
\usepackage{courier}
\usepackage{listings}
\usepackage{xcolor}
\usepackage{color,soul}
\usepackage{amsmath}
\usepackage{multicol}
\usepackage{lipsum}
\usepackage{comment}
\usepackage{amsmath}
\usepackage{subfig}
\usepackage[export]{adjustbox}
\usepackage[inline]{enumitem}
\usepackage{mathtools}
\usepackage{algpseudocode}
\usepackage{tabstackengine}
\usepackage{multirow}
\usepackage{arydshln}
\usepackage{todonotes}
\usepackage{pifont}
\algrenewcommand\textproc{}
\usepackage[all]{nowidow}
\usepackage{array}
\usepackage[skip=4pt,font=normal]{caption}
\usepackage{hyperref}

\usepackage{bm}
\usepackage{soul}
\usepackage{amsthm}
\usepackage{makecell}
\usepackage[export]{adjustbox}
\usepackage{dcolumn}
\usepackage{tabu}

\usepackage{colortbl}

\newcommand{\tbhline}{\noalign{\hrule height 1pt}}

\newcolumntype{|}{!{\vrule width 0.5pt}}
\newcolumntype{?}{!{\vrule width 1pt}}
\newcolumntype{^}{!{\vrule width 1.2pt}}

\renewcommand{\O}[1]{$\mathcal{O}(#1)$}

\algtext*{EndWhile}
\algtext*{EndIf}

\definecolor{myblue}{RGB}{91,155,213}
\definecolor{mydark}{RGB}{0,0,0}

\DeclareRobustCommand{\hllightgray}[1]{{\sethlcolor{lightgray}\hl{#1}}}

\SetKwInOut{Input}{Input}
\SetKwInOut{Output}{Output\,}
\SetKwInOut{Data}{Data}
\SetKwProg{Tree}{Tree}{}{EndTree}

\newcommand{\pluseq}{\mathrel{+}=}

\begin{document}

\title[Performance Portable Back-projection Algorithms]{Performance Portable Back-projection Algorithms on CPUs: Agnostic Data Locality and Vectorization Optimizations}

\author{Peng Chen}
\orcid{1234-5678-9012}
\affiliation{%
  \institution{National Institute of Advanced Industrial Science and Technology}
  \institution{RIKEN CCS}
    \country{Japan}
}
\email{chin.hou@aist.go.jp}

\author{Mohamed Wahib}
\orcid{1234-5678-9012}
\affiliation{%
  \institution{National Institute of Advanced Industrial Science and Technology}
  \institution{RIKEN CCS}
    \country{Japan}
}
\email{mohamed.attia@aist.go.jp}

\author{Xiao Wang}
\orcid{1234-5678-9012}
\affiliation{%
  \institution{Harvard Medical School, Boston}
  \institution{Boston Children's Hospital}
    \country{USA}
}
\email{xiao.wang2@childrens.harvard.edu}

\author{Shinichiro Takizawa}
\orcid{1234-5678-9012}
\affiliation{%
  \institution{National Institute of Advanced Industrial Science and Technology}
    \country{Japan}
}
\email{shinichiro.takizawa@aist.go.jp}

\author{Takahiro Hirofuchi}
\orcid{1234-5678-9012}
\affiliation{%
  \institution{National Institute of Advanced Industrial Science and Technology}
    \country{Japan}
}
\email{t.hirofuchi@aist.go.jp}

\author{Hirotaka Ogawa}
\orcid{1234-5678-9012}
\affiliation{%
  \institution{National Institute of Advanced Industrial Science and Technology}
    \country{Japan}
}
\email{h-ogawa@aist.go.jp}

\author{Satoshi Matsuoka}
\orcid{1234-5678-9012}
\affiliation{%
  \institution{Tokyo Institute of Technology}
  \institution{RIKEN CCS}  
  \country{Japan}
}
\email{matsu@acm.org}

\renewcommand{\shortauthors}{Chen, P. et al.}

\begin{abstract}
Computed Tomography (CT) is a key 3D imaging technology that fundamentally relies on the compute-intense back-projection operation to generate 3D volumes. GPUs are typically used for back-projection in production CT devices. However, with the rise of power-constrained micro-CT devices, and also the emergence of CPUs comparable in performance to GPUs, back-projection for CPUs could become favorable. Unlike GPUs, extracting parallelism for back-projection algorithms on CPUs is complex given that parallelism and locality are not explicitly defined and controlled by the programmer, as is the case when using CUDA for instance. We propose a collection of novel back-projection algorithms that reduce the arithmetic computation, robustly enable vectorization, enforce a regular memory access pattern, and maximize the data locality. We also implement the novel algorithms as efficient back-projection kernels that are performance portable over a wide range of CPUs. Performance evaluation using a variety of CPUs from different vendors and generations demonstrates that our back-projection implementation achieves on average $5.2\times$ speedup over the multi-threaded implementation of the most widely used, and optimized, open library. With a state-of-the-art CPU, we reach performance that rivals top-performing GPUs.
\end{abstract}



\maketitle

\section{Introduction}
Computed Tomography (CT) is a key 3D imaging technology used in several fields such as medical analysis, scientific inspection, and non-intrusive testing. 
Back-projection is a fundamental kernel in most of the image reconstruction algorithms such as FDK~\cite{feldkamp1984practical} and iterative reconstruction algorithms~\cite{geyer2015state,Biguri_2016,biguri2019arbitrarily}
 that employ the algorithms of forward-projection and back-projection,
e.g. MLEM~\cite{shepp1982maximum,hudson1994accelerated}, 
EM~\cite{green1990bayesian}, 
ART~\cite{scarfe2008cone}, and SART~\cite{andersen1984simultaneous}. As a consequence, the back-projection kernel is effectively empowering 100,000s of production CT devices worldwide~\cite{10.1371/journal.pone.0126036}. 
Due to its high computational cost (\O{N^4}~\cite{scherl2012evaluation,lu2016cache}), back-projection is often the compute bottleneck for reconstructing 3D images. That is specially the case in iterative reconstruction algorithms at which back-projection is called repeatedly~\cite{willemink2019evolution}.
To meet the critical demands for rapid image reconstruction, in the past decades, a plethora of accelerators were employed to improve the computational performance of back-projection such as 
linear accelerators~\cite{swindell1983computed},
ASICs~\cite{wu1991asic},
DSPs~\cite{liang2010optimized}, FPGAs~\cite{coric2002parallel,xue2006acceleration,kasik2012advanced},
and distributed systems~\cite{yang2006parallel,10.1145/3295500.3356163,wang2017massively,palenstijn2016distributed,hidayetoglu2020petascale,Gao2019BlockSG}.
Custom processors have been adopted by production CT devices in the earlier generations of CT devices. However, over the last decade, GPUs have increasingly become the main processor used for image reconstruction, due to their programmability and competitive performance~\cite{diez2007implementation, cabral1994accelerated,mueller2007gpus, xu2005accelerating,EKLUND20131073,sabne2017model,rezvani2007ff,zhao2009gpu,zinsser2013systematic,palenstijn2016distributed,biguri2019arbitrarily}.

We argue that there are strong reasons for optimizing back-projection kernels for CPUs, despite GPUs being the de facto processor for back-projection over the last decade. First, there is a noticeable trend of portable and lightweight micro-CT devices~\cite{swain2009state,clark2014micro,compactCT,ying2017micro}. This further makes CT vendors more sensitive to the cost, power, and space requirements that come along with discrete accelerators. For instance, a state-of-art lineup of industrial micro-CT devices (listed in~\cite{xth450}) reports power consumption ranging between 225W to 450W. However, low-end GPUs would consume up to 30\% of that power budget, and high-end GPUs (i.e. Nvidia's P100~\cite{nvidia2016p100} and V100~\cite{nvidia2017corporation}) would consume almost 100\% of the power budget of a micro-CT device~\cite{nvidia2017corporation}.
Second, high-performance CPUs with high bandwidth memory can compete with GPUs in raw performance and performance to power. For example, Fujitsu's A64FX ARM processor~\cite{yoshida2018fujitsu} is equipped with HMB2 memory~\cite{jun2017hbm} and SVE (Scalable Vector Extension)~\cite{stephens2017arm}, which pushes the performance to be comparable to the top of line GPUs in a variety of workloads~\cite{10.5555/3433701.3433763}. Third, integrating accelerators into CT systems increases the system complexity and leads to considerable costs, i.e., hardware and software development. The use of performant CPUs would be preferable since they are a basic component in most CT devices. To conclude, it is crucial to revisit back-projection for CPU, in light of those developments. 

We propose novel optimizations at the algorithm-level, and not at the target CPU hardware level. 
We argue that those carefully designed and novel optimizations at the algorithm level are key to performance portability on a variety of CPUs, having different characteristics. This is in contrast to previous work on CPU back-projection that engineers target-specific optimizations to a single dedicated CPU target~\cite{treibig2013pushing,hofmann2014performance,Xiao:MBIR,karrenberg2012improving,lee2013opencl}. It is important to note that, in the context of performance portability, CPUs have much more divergence in characteristics to optimize for, in comparison with GPUs. For instance, the vector width changes from generation to generation (for even the same vendor), while in Nvidia GPUs, the unit of execution (CUDA warp~\cite{cudaToolkit}) maintained the same size in all generations.

We briefly list here the algorithm-level optimizations for back-projection we propose for CPUs. First, effective data locality and regular memory access patterns are critical to improving the efficiency of back-projection kernels. To improve the data locality~\cite{stallings2003computer}, we introduce a generic scheme to reschedule the loop ordering, employ data blocking, and change the data layout rearrangement. This is in contrast to other non-portable methods that rely on the gather load intrinsic to deal with the irregular memory access~\cite{treibig2013pushing}. Second, assuring that the SIMD vector units of the CPUs are fully utilized is crucial for performance. Designing back-projection algorithms that are fully vectorizable by compilers, yet being performance portable across CPUs from different vendors is necessary, yet challenging. To assure a loop could vectorize automatically, it is necessary to simplify the iteration space and data accesses such that it can be auto-vectorized by the compiler. 
We hence gear our optimization techniques to enable the compiler to consistently vectorize the back-projection kernel, regardless of the vector width. Third, back-projection requires high computational resources. It is crucial to improve its performance at the algorithmic level by reducing the arithmetic computation and memory access to the least possible. To that end, we propose a novel double-buffer sub-line algorithm for efficient interpolation at the sub-pixel precision. We also reduce the number of arithmetic operations, by exploiting geometric symmetry, while maintaining contiguous memory access. In addition, we block the sub-line interpolation, to improve the locality without any impediment to the compiler's capability of automated vectorization.

The contributions in this paper are:
{
{
    \setlength{\leftmargini}{15 pt}
    \begin{itemize}
        \item We propose a collection of novel algorithm-level optimizations for back-projection on CPUs. Our optimizations reduce the computational cost of the projection operations, improve data locality, and consistently enable auto-vectorization. 
        \item We provide a performance portable implementation: a single OpenCL source for all CPUs supporting OpenCL and a single OpenMP source for CPUs that do not support OpenCL (without any trace of processor-specific intrinsics or optimizations). 
        \item We demonstrate that our kernels achieve on average $5.2\times$ speedup over the most widely used and optimized multi-threaded library, on a variety of CPUs from different vendors and generations. Also, we achieve better performance on the Fujitsu A64FX than commonly used open-source CUDA implementation of back-projection on Nvidia V100 Volta GPU (when accounting for unavoidable data movement overhead between host and device).
    \end{itemize}

}
}
The rest of this paper is organized as follows:
In Section~\ref{sec:background}, we introduce the background.
Section~\ref{sec:proposed algorithm} illustrates the proposed algorithms.
Section~\ref{sec:evaluation} shows the evaluated result.
In Section~\ref{sec:related work}, we elaborate on the related work.
Finally, Section~\ref{sec:conclusion} concludes.

\section{Background}\label{sec:background}
In this section, we introduce the details of image reconstruction using the back-projection algorithm and describe the basics of OpenCL for multicore CPUs. 
\subsection{CT Image reconstruction}\label{sec:fdk}
This section illustrates the 3D image reconstruction algorithm for Cone-Beam Computed Tomography (CBCT) as presented by Feldkamp et al.~\cite{feldkamp1984practical}, including the CBCT geometry and back-projection algorithm.

\subsubsection{\bf{Geometry of CT system}}\label{sec:ct-geometry} 
\begin{figure}[t]
  \begin{center}
    \includegraphics[clip,width=0.495\textwidth]{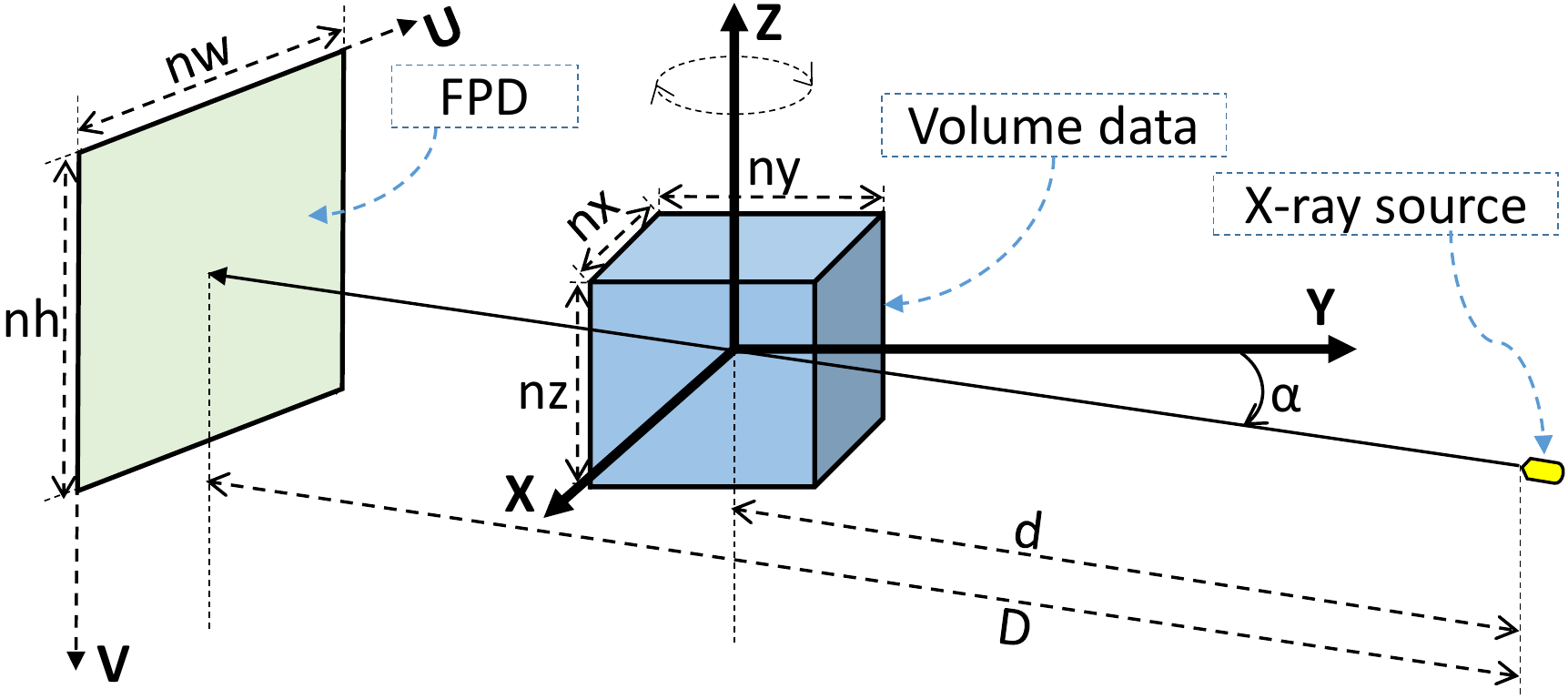}
    \caption{The triangular geometry of a CT system.}
    \label{fig:geometry}
  \end{center}
\end{figure}

Figure~\ref{fig:geometry} shows the triangular geometry of CBCT system~\cite{scarfe2008cone}. The X-ray source is some form of a microfocus X-ray tube. The Flat Panel Detector (FPD) is a digital radiography imaging sensor, similar to that of digital photography. The distances of the source to the rotation axis (the Z-axis) and FPD are \emph{d} and \emph{D}, respectively. The width and height dimensions of the FPD, in the unit of pixel, are {\bf{nw}} and {\bf{nh}}, respectively. Note that the V-axis of FPD is parallel to Z-axis.
The sizes of 3D volume data in a unit of voxel are {\bf{nx}}, {\bf{ny}}, and {\bf{nz}}. The default data layout of volume data is row-major order. As a typical pinhole model~\cite{hartley2003multiple}, all geometric information can be presented as a matrix of size 3$\times$4 (called {\emph{projection matrix}}), which is used for back-projection computation, e.g. projecting a voxel to the plane of FPD as \textit{mat} in Listing~\ref{alg:bp-baseline}. For simplification, we ignore the computation of obtaining the projection matrix via the geometry information. The detailed formulation is available at~\cite{wiesent2000enhanced}. Computing the projection, i.e., mapping a 3D point (i, j, k) to the plane of FPD (the UV plane) is shown in Listing~\ref{alg:bp-baseline} (lines 7$\sim$11).

\input{listings/alg-bp-baseline}
\input{listings/alg-dot4}

\subsubsection{\bf{Back-projection}}\label{sec:BP}
The computational complexity of back-projection is \O{N^4}~\cite{scherl2012evaluation,lu2016cache}. Listing~\ref{alg:bp-baseline} shows the pseudocode of the reference multi-threaded back-projection algorithm as implemented in the widely used RTK library\footnote{Reconstruction Toolkit (RTK): https://www.openrtk.org}. Note that the pseudocode also matches the multi-threaded implementation of the other most widely used library, RabbitCT~\cite{rohkohl2009rabbitct}. A parallel loop (line \emph{4}) is where the work is divided among the threads to take advantage of the multi-cores of the processor.
The inputs to the back projection kernel are: the 2D filtered projections {\bf{img}}, the 3D output volume of the filtering computation {\bf{volume}}, and the dimensions expressed as (np, nh, nw). Throughout this paper, {\bf{np}} refers to the total number of projections, and {\bf{nb}} refers to the number of projections in a single batch. More specifically, the projections are equally split into disjoint subsets of projections, where each subset includes a number of projections \emph{nb} that is used to update the volume data as a batch.

The projection matrix {\bf{mat}} is used to project the voxel to the FPD plane. The projected coordinate is written as \emph{x} and \emph{y} in lines \emph{10} and \emph{11}. The value of \emph{z} (line \emph{8}) is used to derive the projection coordinate, as well as a weighting factor to update the volume data in lines \emph{13$\sim$14}.
The {\bf{dot4}} operator (line \emph{10}), is a customized operator for the inner product as in Listing~\ref{alg:subpixel} that is called three times in the most inner loop. As listing~\ref{alg:subpixel} shows, a bilinear interpolation function, namely {\bf{Bilinear\_Interpolate}}, is used to fetch the intensity value of the 2D projections.

\subsection{Parallel Computation by OpenCL}\label{sec:opencl-on-cpu}
OpenCL provides a portable way to take advantage of the SIMD-accelerated processors with a programming approach of single instruction multiple threads. The OpenCL kernels can be compiled dynamically at runtime for the target architecture. OpenCL abstracts an open standard platform for parallel programming and provides a uniform programming interface that is often used to write portable and efficient code for a variety of processors, e.g. CPUs, GPUs, and FPGAs. 
As an open standard defined by Khronos Group~\cite{wiki:Khronos_Group}, OpenCL uses a host/device programming model.
Programming by OpenCL involves writing host code, often in C/C++, that runs on the host, and a kernel code developed in C, that runs on the device or accelerator.
The host codes are often compiled by a general C/C++ compiler (e.g. gcc/g++), and the kernel codes are built at runtime by a target-specific compiler.
The computing unit in OpenCL is a work-item. A work-group is a collection of work-items abstracted as a multi-dimensional descriptor named \emph{NDRange}. Note that the work-items within the same work-group can exchange data using the local memory.

\subsection{Terminology}\label{sec:terminology}
We define the image reconstruction problem and performance metrics in this section.
{
The image reconstruction problem is defined as {\bf{$nw{\times}nh{\times}np{\Rightarrow}nx{\times}ny{\times}nz$}} , where $nw{\times}nh$ and $np$ are the size and number of input projections, respectively. The size of the output volume is defined as $nx{\times}ny{\times}nz$. 
The performance metric we use is calculated as
${\bf{nx*ny*nz*np*t^{-1}*10^{-9}}}$,
where \emph{t} denotes the run-time in a unit of second.
The performance unit of the kernel commonly used in image reconstruction algorithms is {\bf{GUPS}}, which stands for Giga Updates per Second.

}

\section{Proposed Back-projection Algorithms}\label{sec:proposed algorithm}
In this section, we discuss a variety of algorithms for back-projection that employ different optimization schemes corresponding to different CPU features, without loss of generality with regard to CPU vendor and generation. The optimization schemes do not interfere with each other and could be combined.
We improve the data locality and memory access pattern by transposing projections and volume data. To fully utilize the multi-cores and vector units, we also take advantage of both OpenMP and OpenCL to speed up the back-projection computation. Auto-vectorization can be performed at compile time when using either OpenMP or OpenCL. Without explicit vector instructions, and by carefully writing the code, the OpenCL compiler~\cite{intel:opencl} can automatically vectorize instructions by packing multiple work-items together and running them in a SIMD fashion. To improve the data locality and reduce the arithmetic computation, we use a novel sub-line algorithm to cache the interpolated data at sub-pixel precision. More specifically, we improve the data locality of the bilinear interpolation while enabling auto-vectorization.

\subsection{Algorithmic Optimizations}\label{sec:opt-bp}
This section describes the algorithmic changes we introduce to improve the memory access pattern, reduce arithmetic computation, and expose data-parallel operations. These optimizations apply to both single-core and multicore. Those optimizations at the algorithmic level are key to performance portability.

As shown in Listing~\ref{alg:bp-baseline}, the back-projection is driven by voxels, which are updated by their mapped pixels at projections. The values of voxels are independently updated, i.e. an embarrassingly parallel point-wise operation. Hence, the algorithmic optimizations introduced should, in theory, scale with the capacity of computing resources, i.e. number of cores. 
The following section elaborates on the algorithmic optimizations.

\begin{figure}[t]
  \begin{center}
    \includegraphics[clip,width=0.495\textwidth]{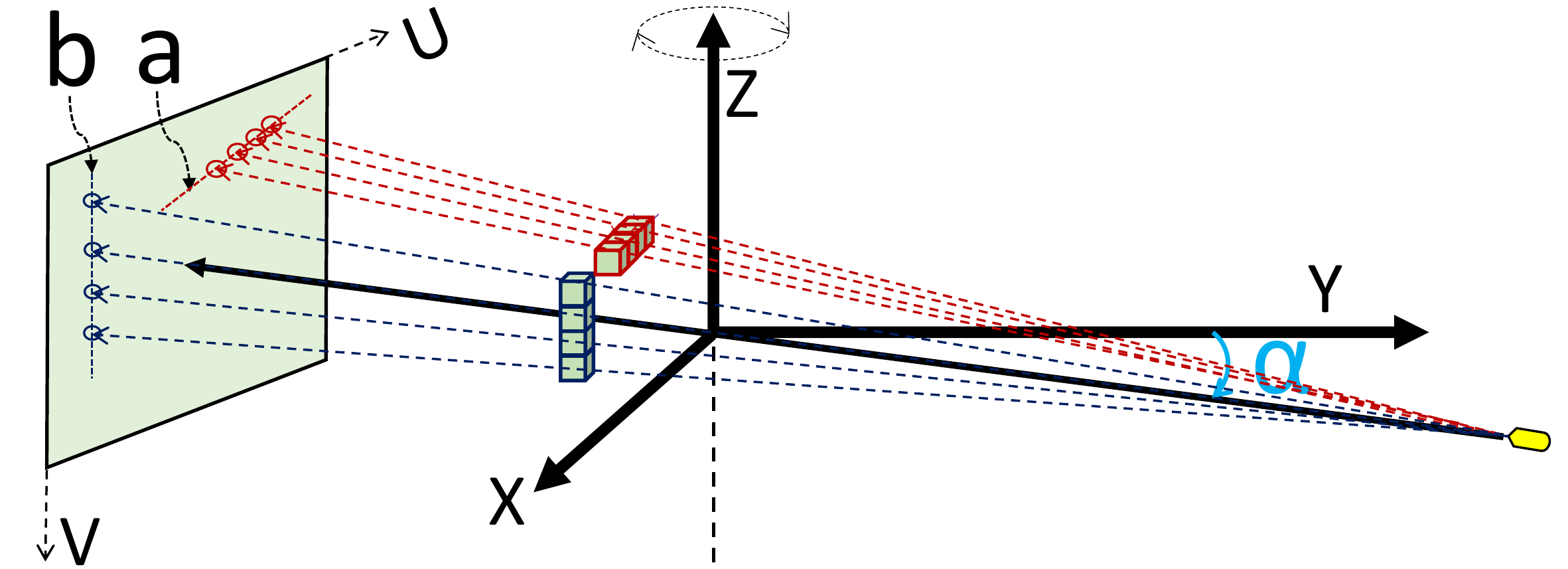}
    \caption{The motivation for transposing projections and volume data. The red and blue voxels are parallel to the X-axis and Z-axis, such that their projection points on the UV plane are lines $\bar{a}$ and $\bar{b}$, respectively. The slope of line $\bar{a}$ changes according to the rotation angle $\alpha$. The line $\bar{b}$ is invariably vertical (namely parallel to both V-axis and Z-axis).}
    \label{fig:geometry-motivation}
  \end{center}
\end{figure}
\input{algorithm/alg-scratchpad-subline}

\input{figures/alg-subline}

\subsubsection{\bf{Towards a Regular Memory Access Pattern}}\label{sec:improve-data-layout}
As Listing~\ref{alg:bp-baseline} shows, each voxel is updated by the intensity of projections, according to the projected position at sub-pixel precision using the bilinear interpolation scheme in Listing~\ref{alg:subpixel}.
Figure~\ref{fig:geometry-motivation} can be used to visualize the mechanism of bilinear interpolation and also illustrates the optimization of back-projection by transposing the projection and volume data. As shown in the figure, the memory access for the bilinear interpolation along the line $\bar{a}$ is irregular (sloped access to a 2D mesh). Furthermore, the slope of line $\bar{a}$ changes with the rotation angle of $\alpha$. 
However, the projections of blue voxels (parallel to Z-axis) are in a line $\bar{b}$, which is invariably parallel to V-axis and Z-axis. Therefore, the memory access along line $\bar{b}$ would be regular and unit-strided, if we transpose the projections.

The proposed Algorithm~\ref{alg:scratchpad-subline} shows the optimization using transposed projections and volume data.
More specifically, we optimize the data access pattern by transposing the two-dimensional projections \emph{img} (input argument). Note that the transposing operation requires marginal time in comparison to the back-projection and can be performed immediately after the pre-processing filtering step, and before the back-projection starts.
We also reorganize the loop that iterates the voxels along the Z-axis in the original volume to get unit-strided access over the transposed volume data as the input argument \emph{volume} shows. 
More importantly, this transposition operation builds a foundation for further optimizations such as exploiting symmetry and improving the bilinear interpolation scheme to be vectorizable.

\subsubsection{\bf{Reducing Arithmetic Computation}}\label{sec:reduc-computation}
We reduce the volume of the arithmetic computation by:
a) capitalizing on the loop reordering introduced with transposition by moving the loop invariant compute portions outside the inner loop (i.e. hoisting), and b) exploiting geometric symmetry.
The data reuse is shown in lines 4$\sim$7 of Algorithm~\ref{alg:scratchpad-subline}. 
The intermediate variables (namely \emph{F}, \emph{W}, and \emph{X}) are only computed once and their results are shared by the inner loops for further computations, e.g. projection operation, weighting projection intensity, and updating the volume data.
Note that the variables \emph{F}, \emph{W}, and \emph{X} contain a small set of values of \emph{f}, \emph{w}, and \emph{x}, respectively (as shown in Listing~\ref{alg:bp-baseline}).
In this optimization, we exploit the geometric characteristics of back-projection where the values of \emph{z} and \emph{x} are constant values when \emph{i}, \emph{j}, and the projection matrix (with projection angle $\theta$) are fixed values. That is since the value of \emph{z} is equal to $d-xcos{\theta}-ycin{\theta}$ (as introduced in~\cite{avinash1988principles, lu2016cache}), where \emph{x} and \emph{y} are coordinates of voxels that correspond to the indexes of \emph{i} and \emph{j}. The values of \emph{x} are also independent of \emph{i} and \emph{j}. That is since the projections of voxels are parallel to the V-axis (or Z-axis in Figure~\ref{fig:geometry}) at the FPD plane, as introduced in Section~\ref{sec:ct-geometry}.

We also reduce the arithmetic computation by exploiting the geometric symmetry proposed by Zhao et al.~\cite{zhao2009gpu}.
As Figure~\ref{fig:geometry} shows, for two voxels that are symmetric to the XY plane, their projections at FPD (namely UV plane) are symmetric to the horizontal centerline of FPD.  As line 13 of Algorithm~\ref{alg:scratchpad-subline} shows, we exploit this symmetry by only performing half of the projection computations for \emph{y}; the symmetric positions of \emph{y} can be derived using a single instruction rather than using complex dot and multiply operations.  

More specifically, the original number of dot operations is $3*np*nx*ny*nz$, the optimized number of operations is reduced to $np*nx*ny*(1+nz/2)$.
Hence, the ratio of reduced computations for dot operations may be written as $({5-2/nz})/{6}{\approx}{5}/{6}$ since $nz\gg2$ in real-world applications.

\input{listings/alg-ocl-bp-symmetry-lm}

\subsubsection{\bf{Improving locality \& Reducing Memory Accesses}}\label{sec:reduc-memory-access}
We present the details of reducing memory accesses for improving back-projection performance in this section. As discussed by Treibig et al.~\cite{treibig2013pushing} and Hofmann et al.~\cite{hofmann2014performance}, back-projection is a bandwidth-limited algorithm and the performance bottleneck is the voxel updates (see line $14$ of Listing~\ref{alg:bp-baseline}). In addition to the coalesced access to the memory of volume data proposed by Treibig et al.~\cite{treibig2013pushing}, we improve the algorithm by further reducing the memory accesses to volume data. In the following paragraphs we elaborate.

First, existing multi-threaded implementations (shown in Listing~\ref{alg:bp-baseline}) do not split the projections into subsets. This is since the interpolation schemes used in those implementations would reduce the locality if accessing multiple images in the same batch. Contrarily, using batches of images to update the volume is preferable since that would reduce the writes to memory. To use batched projections, one would require a new interpolation scheme to address this issue (more on that in Section~\ref{sec:sub-line}).

Second, we reduce memory access to the volume data by using registers to accumulate the partial results (as shown in Algorithm~\ref{alg:scratchpad-subline}). In line 14$\sim$27, we use two registers to accumulate the partial results of voxels for multiple projections in batched processing, then update the volume data by the values accumulated in the registers. Since the cost of accessing registers is negligible with regard to memory access, this batched processing for projections and accumulation of partial results can significantly reduce the number of memory accesses. The original number of elements accessed for the volume data is $4*np*nx*ny*nz$, while after the optimization the number of elements accessed drops to $4*np*nx*ny*nz/nb$. Therefore, in comparison to the baseline implementation, the ratio of memory accesses to the volume data becomes $1/nb$. With a larger value of \emph{nb}, we can move the bandwidth bottleneck of the memory access from volume data to projections. More specifically, the performance of the proposed algorithm may be bounded by the memory accesses to the projections for the bilinear interpolations. In Section~\ref{sec:evaluation}, we demonstrate the performance gain of this optimization.

Finally, we also propose a novel bilinear interpolation with vectorization. This optimization can also reduce the memory accesses to some extent as will be discussed in later sections.

\subsubsection{\bf{Vectorizating Operations}}\label{sec:vectorization}
In this section, we present the details of how to enable full utilization of the vector units in the projection operation and interpolation computation in back-projection.
Regarding the projection computation (e.g. the computation of \emph{x}, \emph{y}, and \emph{z} in~Listing~\ref{alg:bp-baseline}), each inner product is performed on two vectors of size $1\times4$ at single precision. This inner product operation can be perfectly performed by the vector unit, e.g. the built-in function \emph{dot} is employed for this computation in our OpenCL-based implementation (as will be elaborated later). Note that we benefit from the well-aligned projection matrix of size 3$\times$4 in back-projection algorithms. This is in contrast to other back-projection algorithms, such as in~\cite{lu2016cache}, at which the projection computation is not vectorizable due to the irregular memory access and compute pattern.

Since we use transposed projections and volume data, the linear interpolation operation can perform the memory accesses in a regular pattern and its computation can also be well vectorized with wide SIMD registers. In Algorithm~\ref{alg:scratchpad-subline}, we can use vector registers for the accumulation of the variables \emph{sum} and \emph{\_sum}. 

In addition, the computation of the second linear interpolation, as in Figure~\ref{fig:use-subline}, can be vectorized. More specifically, the memory accesses to update the volume data (lines \emph{26} and \emph{27} in Algorithm~\ref{alg:scratchpad-subline}) can be performed in a vectorized fashion. This vectorized operation can improve the effective memory bandwidth for back-projection, especially in CPUs with High Bandwidth Memory (HBM), e.g. Fujitsu A64FX processor.

\subsubsection{\bf{A Novel Bilinear Interpolation Scheme}}\label{sec:sub-line}
This section discusses our scheme for bilinear interpolation. In linear interpolation, we calculate the target element as a weighted sum from two contributors. In this paper, we use the OpenCL built-in, vectorizable, \emph{mix} linear blending function~\cite{munshi2009opencl}.
Bilinear interpolation performs a two-dimensional interpolation on a rectilinear grid and thus, it estimates an output element with four known contributors. The conventional back-projection algorithm can not perform the compute pattern in a regular access pattern: bilinear interpolation must be processed on a per-point basis since the data points of the projections are distributed on a sloped line. We capitalize on the transposition of the projections and volume to introduce a bilinear interpolation scheme that blends two lines of pixels in the vertical direction, followed by interpolating in the horizontal direction, where both have a regular access pattern and could be vectorized.

We elaborate on the effect of the proposed scheme on the vectorization of operations: both memory accesses and arithmetic computations. In Figure~\ref{fig:subline}, the variables \emph{ptr0} and \emph{ptr1} represent the addresses of two neighboring lines in a transposed projection. 
We load the data from the memory in a coalesced pattern and thus, the wide SIMD intrinsic can be employed as in the code example in Figure~\ref{fig:subline-by-opencl}.
Figure~\ref{fig:subline-by-c} shows an alternative vectorized OpenMP implementation. 
It is important to note the regular access pattern for the memory access and arithmetic computations, i.e. loading data from the memory of ptr0/ptr1, storing the result to the memory of sMem, and the linear interpolation operations by the mix function. More specifically, we benefit from this scheme as follows: first, we linearly blend two lines of projections (as ptr0 and ptr1 in Figure~\ref{fig:subline-algorithm}) into a line of data where all elements in these two lines are only accessed once in a regular pattern. Second, either of the linear interpolations in the horizontal or vertical direction could be easily parallelized. Third, the second linear interpolation can be performed via cache-optimized access of the intermediate memory buffer between the vertical and horizontal steps (\emph{sMem} in line 12 in Algorithm~\ref{alg:scratchpad-subline}) with better locality and fewer linear interpolation operations.

\input{tables/tbl-ComputeDevice}

\subsection{OpenCL-optimized Back-projection}
This section discusses how to take advantage of OpenCL features to implement a performance-portable collection of back-projection kernels based on the optimization techniques we introduced in previous sections.
\subsubsection{\bf{OpenCL kernel implementation}}
In Listing~\ref{alg:ocl-bp-symmetry-lm}, we present a performance-portable OpenCL implementation of the back-projection kernel that includes all the discussed optimization techniques in Algorithm~\ref{alg:scratchpad-subline}).

We elaborate on optimizations techniques Listing~\ref{alg:ocl-bp-symmetry-lm} as follows:
{
\begin{enumerate*}[label=(\Roman*)]
\item we use constant memory~\cite{gaster2012heterogeneous} to cache the projection matrix \emph{mat}. The size of each projection matrix is as small as 48B (sizeof(float)*4*3)
\item we pre-pack the indices of $i$ and $j$ in volume data to an array (argument \emph{vecIJ} to \emph{bp\_optimized} kernel). Each work-group processes all voxels in a single vertical line and thus the indices of \emph{i} and \emph{j} can be shared by all work-items in a work-group. We use local memory to share the indices as the variable \emph{ij} shows. 
\item We use local memory to cache the read-only shared data \emph{F}, \emph{X}, and \emph{W} (line \emph{6}). Note that the batch number (namely \emph{nb}) ranges between 1$\sim$32 in our implementations. Hence, the required size of local memory is $sizeof(float)*B*3$, which is within the capacity of local memory. All of these values are only computed once as in lines 11$\sim$16 and reused by all work-items in a work-group. Finally, a barrier is required, line \emph{17}, to synchronize all work-items in a work-group. 
\item Using the proposed bilinear scheme (as in Section~\ref{sec:sub-line}), we perform the linear interpolation via fast local memory (lines 23$\sim$26).
More importantly, we can vectorize the computation as Figure~\ref{fig:subline-algorithm-code} shows. 
\item By using the pixel values in local memory (namely \emph{sMem}), the second linear interpolation can be performed as shown in Figure~\ref{fig:use-subline} (line \emph{34}). 
\item This kernel also exploits the geometric symmetry to reduce the computation by reusing values computed at the symmetric point, i.e. the value is computed once to be used by the point and its symmetric point (shown as the variables \emph{y} and \emph{\_y}).
\end{enumerate*}
}
We use two-dimensional NDRange to launch this kernel, the local and global parameters may be written as (nz/2, 1) and (nz/2, sizeIJ), respectively. Note that sizeIJ is the number of elements in the array vecIJ. We emphasize the intrinsic that requires the id of work-items in bold font, e.g. get\_global\_id, get\_local\_size \dots, etc. Several built-in functions are used in our implementation such as dot, mix, and convert\_T. For more explanations on these built-in functions, the reader can refer to literature as in~\cite{munshi2011opencl}.

\input{algorithm/alg-prefetch}
\subsubsection{\bf{Prefetching}}\label{sec:prefetch}
The use of OpenCL allows us to take advantage of the double buffers technique to overlap the loading operation and computation. As shown in Algorithm~\ref{alg:prefetch}, we declare dual local memory buffers, which are emphasized in gray color. Hence, we can perform the load operation (line \emph{6}) and computation (line \emph{8}) using different buffers. However, the allocated memory is doubled, and thus, the available size of processing projections will be limited by the capacity of local memory.

\section{Evaluation}\label{sec:evaluation}
In this section, we introduce the evaluation environment, report the results of our experiments, and discuss the advantages and limitations of the proposed algorithms.
\subsection{Experiments Setup}
Table~\ref{tbl:evaluation-env} shows the CPUs we use to evaluate our implementation. We use the same OpenCL implementation on all CPUs, without any customization or optimization to specific targets. For CPUs for which we did not have access to vendor support of OpenCL, we use the same OpenMP implementation, namely the ARM and AMD processors in Table~\ref{tbl:evaluation-env}. 

For the CPUs that use OpenMP library, we employ GCC 9.1.0 (with OpenMP 4.5) for compiling all kernel versions, and "-O3 -fopenmp -lpthread -std=c++11 -march=native -ftree-vectorize -ftree-slp-vectorize" compilation options. For A64FX processor we use Fujitsu's compiler FCC 4.0.0 with the compiler options
"-Kopenmp -Kfast -Ksimd=auto -Kassume=memory\_bandwidth -O3 -Nlibomp". 
On the CPUs that use OpenCL, i.e. Intel CPUs, we use OpenCL SDK-2019.5.345 and runtime 18.1.0 for developing and running OpenCL kernels.
All OpenCL kernels are compiled with the compiler options "-cl-mad-enable -cl-fast-relaxed-math". Nvidia Tesla P100/V100 GPUs and CUDA 10.0 SDK~\cite{cudaToolkit} are also used for performance comparison.

In Table~\ref{tbl:kernel-names}, we list a collection of back-projection kernels that apply the optimizations we proposed in Section~\ref{sec:proposed algorithm}. We take advantage of both OpenMP and OpenCL to optimize back-projection. The kernel variants of the employed optimization techniques are named "Transpose", "Share", $\dots$, and "Prefetching" and they correspond to the optimization techniques discussed in Section~\ref{sec:opt-bp}. A collective combination of all optimization techniques (namely \emph{symmetry\_pf\_cl} in Table~\ref{tbl:kernel-names}) can be found in both Algorithm~\ref{alg:scratchpad-subline} and Listing~\ref{alg:ocl-bp-symmetry-lm} for the OpenCL version.

\input{tables/tbl-kernelNames}

\input{tables/tbl-problems}

\begin{figure}[t]
  \begin{center}
    \includegraphics[clip,width=0.470\textwidth]{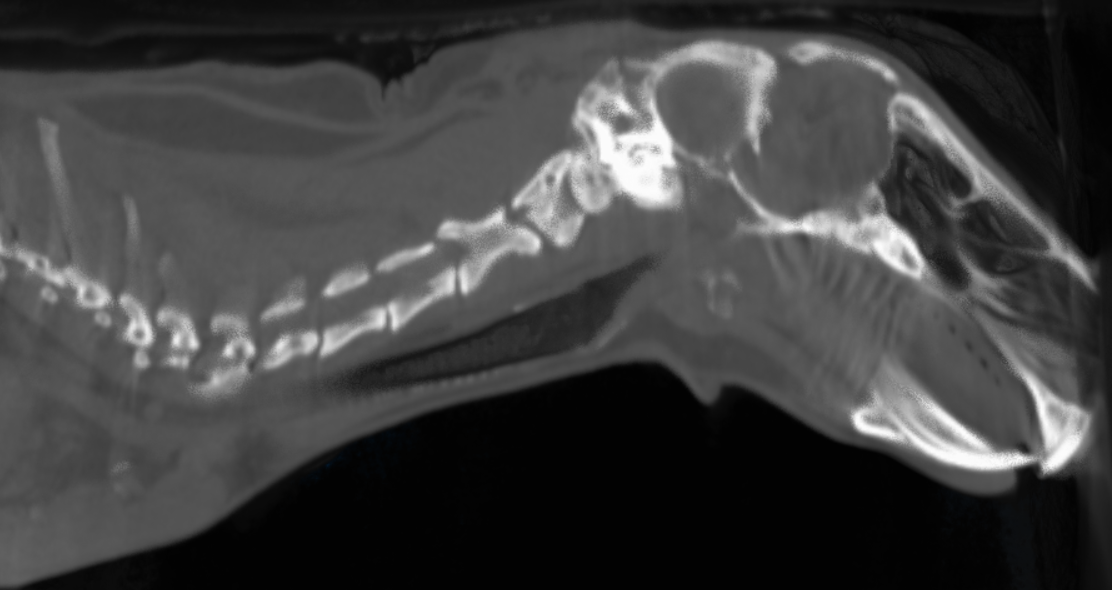}
    \caption{Using our algorithm to reconstruct the volume of the RabbitCT dataset~\cite{rohkohl2009rabbitct} ($512\times512\times512\Rightarrow512^3$). Rendered by ImageJ 3D Viewer~\cite{rasband1997imagej}.}
    \label{fig:rabbitCT}
  \end{center}
\end{figure}

All evaluations are conducted at single precision. In Table~\ref{tbl:problems}, we list the image reconstruction problem sizes, which are labeled as {{\bf}P1, P2,$\dots$, P10}.  The sizes of the experimental projections include $256^2$, $512^2$, and $1024^2$; the number of projections is fixed as 512; the output problems include $256^3$, $512^3$, $1024^3$, and $1300^3$. It is noteworthy that the required memory capacity to store a volume of size $1300^3$ is $\sim$8.2 GB.

\subsection{Image Reconstruction Results}
In Figure~\ref{fig:rabbitCT}, we show an example of the generated volume data by our algorithms. The projections are from a real-world CT scanner described in the RabbitCT dataset~\cite{rohkohl2009rabbitct}.
Since the arithmetic computation is independent of the content of projections and volume data, we also use the volume data of RabbitCT to generate a variety of projections as described in Table~\ref{tbl:problems} by a forward-projection tool in the RTK.
To verify the output, we compare the reconstructed volume data with the results by RTK library, the Root Mean Square Error~\cite{wiki:Root-mean-square_deviation} threshold is less than 10e-5.
Additionally, we employ the ImageJ~\cite{rasband1997imagej} (an image processing tool) to render each generated volume data and manually inspect them.

\begin{figure}[t]
\centering
\subfloat[Dual E-2630 CPUs.] {
 \includegraphics[width=0.24\textwidth]{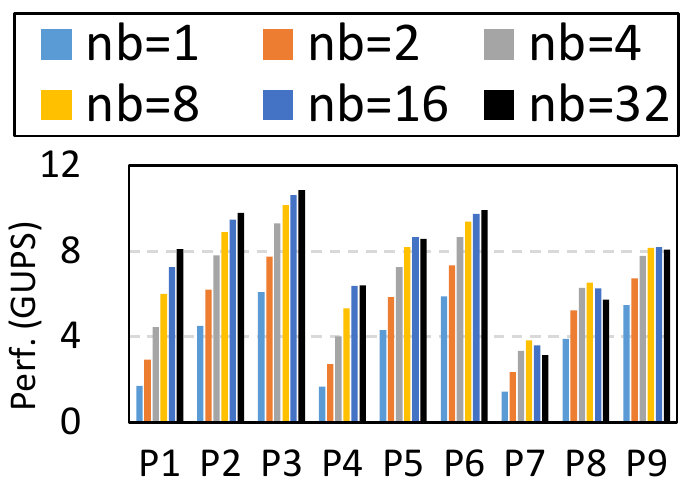}
 \label{fig:2630}
 }
 \subfloat[Dual E-2650 CPUs.] {
 \includegraphics[width=0.24\textwidth]{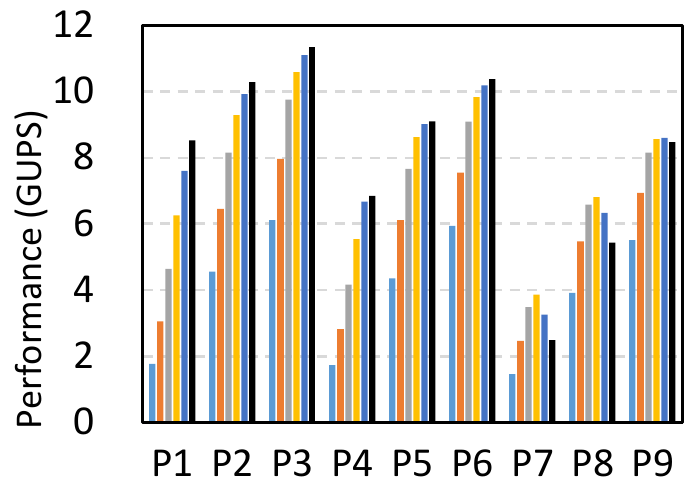}
 \label{fig:2650}
 }
 \vspace{-0.5\baselineskip}
 \subfloat[Dual Gold 6140 CPUs.] {
 \includegraphics[width=0.24\textwidth]{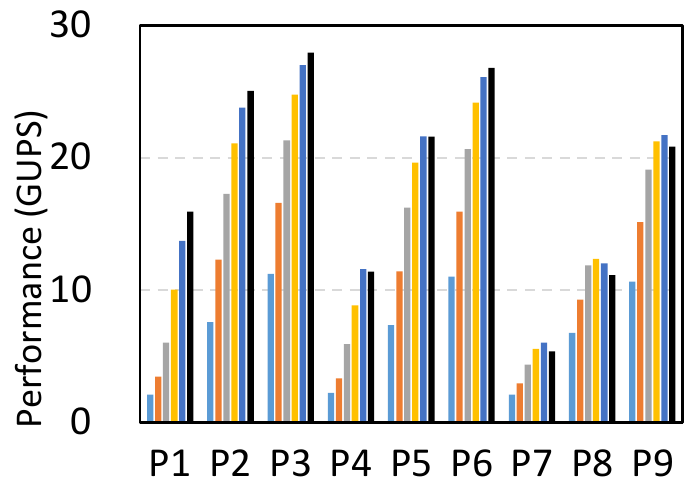}
 \label{fig:gold}
 }
 \subfloat[A single i7-9700K CPU.] {
 \includegraphics[width=0.24\textwidth]{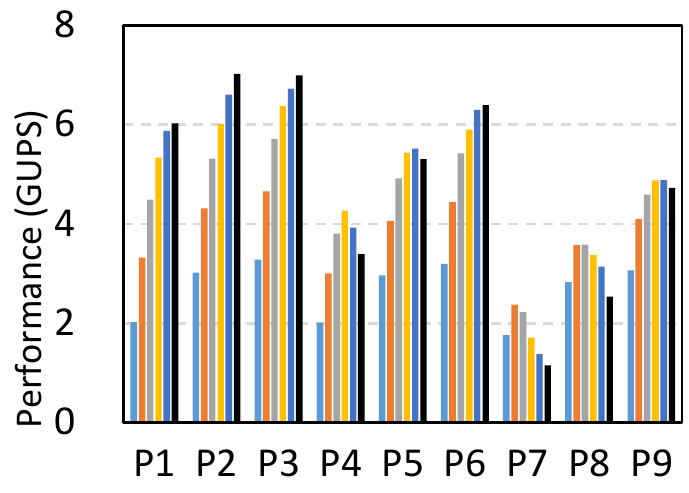}
 \label{fig:9700k}
 }
\caption{Performance of top performing kernel (\emph{symmetry\_pf\_cl}) with different batch numbers (\emph{nb}).
}
\label{fig:batch-number}
\end{figure}

\begin{figure*}[t]
\centering
\subfloat[Dual E-2630 CPUs.] {
 \includegraphics[width=0.328\textwidth]{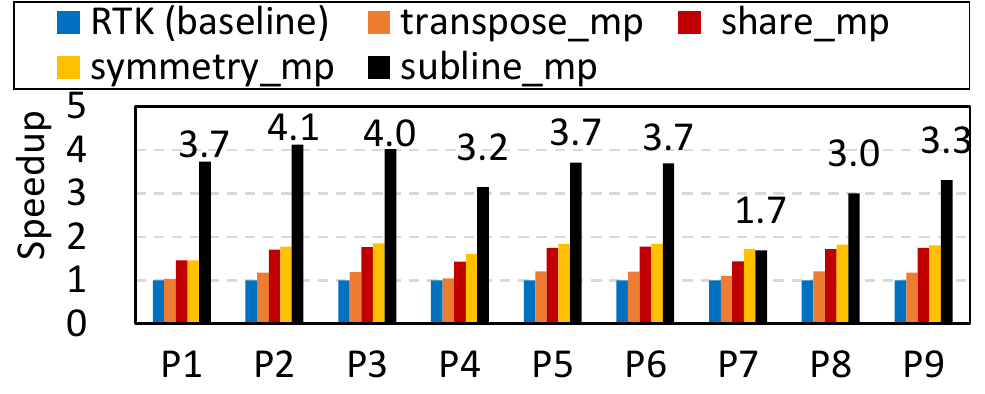}
 \label{fig:2630}
 }
 \subfloat[Dual Gold-6140 CPUs.] {
 \includegraphics[width=0.328\textwidth]{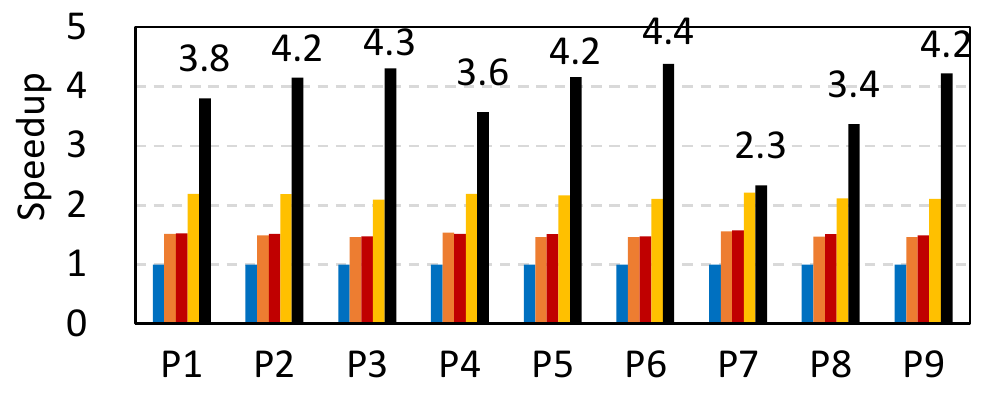}
 \label{fig:2630}
 }
 \subfloat[A single i7-9700K CPU.] {
 \includegraphics[width=0.328\textwidth]{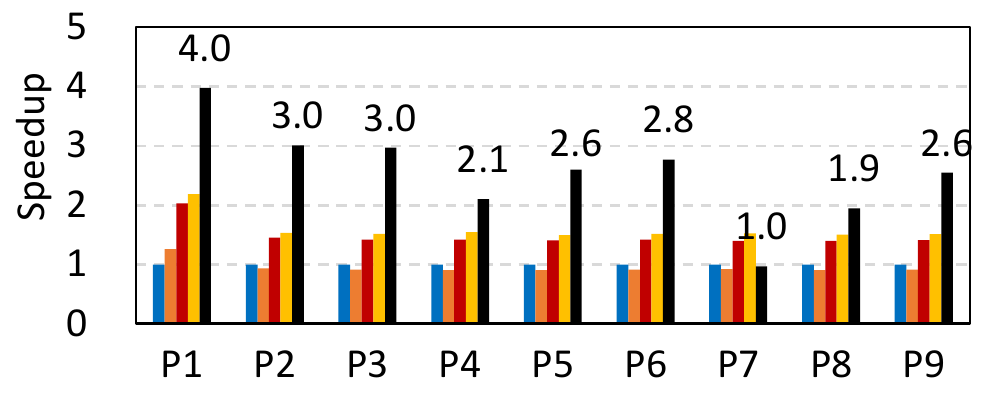}
 \label{fig:2630}
 }
 \vspace{-0.5\baselineskip}
 \subfloat[Dual Epyc-7452 CPUs.] {
 \includegraphics[width=0.328\textwidth]{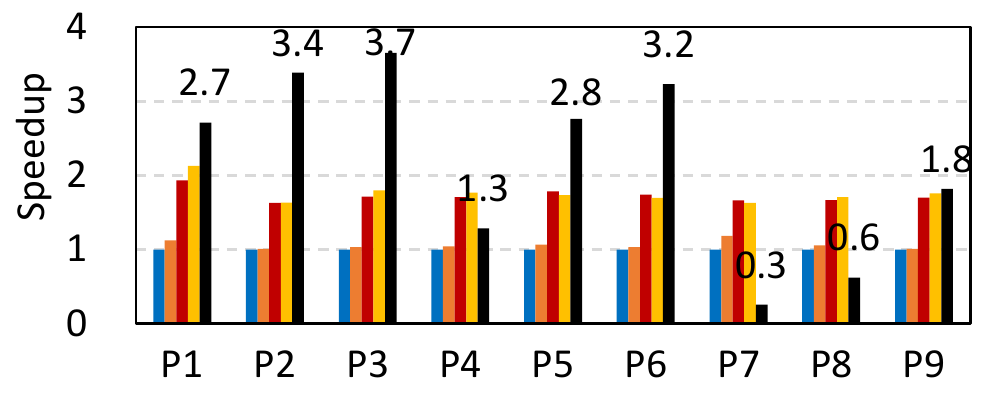}
 \label{fig:2630}
 }
\subfloat[Dual ARM-CN9975 CPUs.] {
 \includegraphics[width=0.328\textwidth]{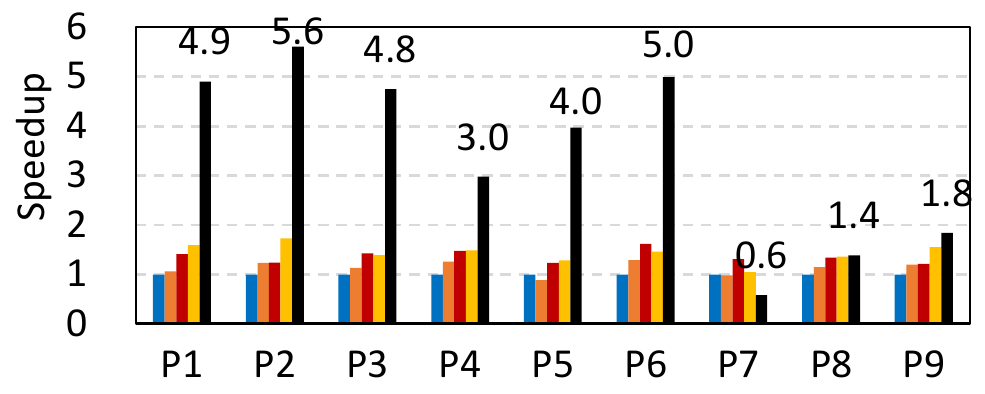}
 \label{fig:2630}
 }
 \subfloat[A single A64FX CPU.] {
 \includegraphics[width=0.3285\textwidth]{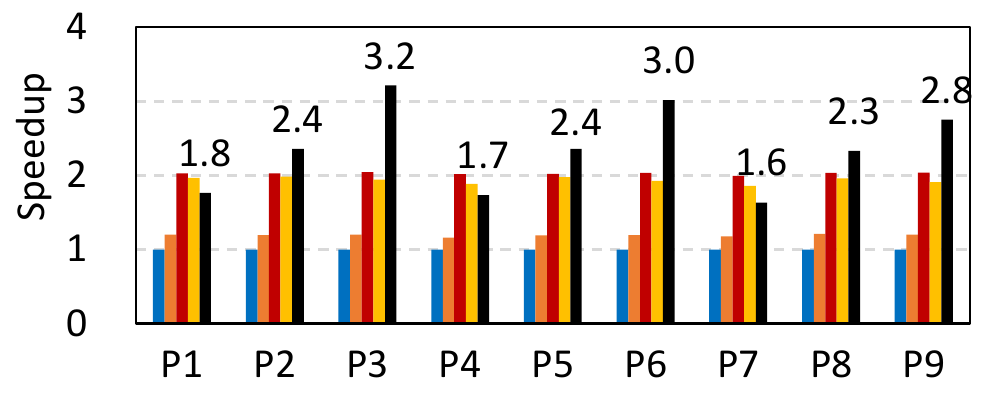}
 \label{fig:2630}
 }
\caption{Speed up for OpenMP implementation for different levels of optimizations.}
\label{fig:batch-speedup}
\end{figure*}

\begin{figure*}[t]
\centering
\subfloat[Dual E-2630 CPUs.] {
 \includegraphics[width=0.5\textwidth]{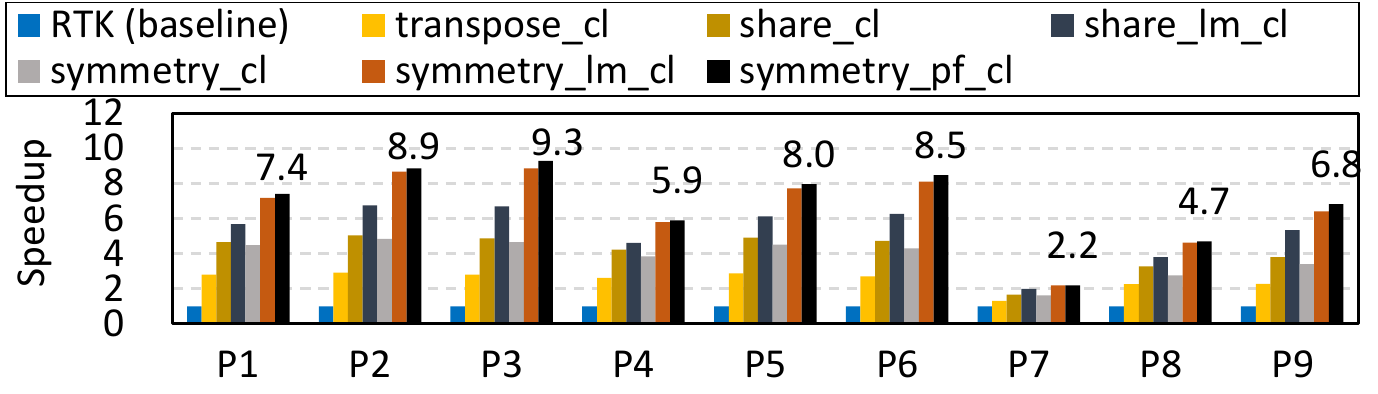}
 \label{fig:2630}
 }
\subfloat[Dual E-2650 CPUs.] {
 \includegraphics[width=0.5\textwidth]{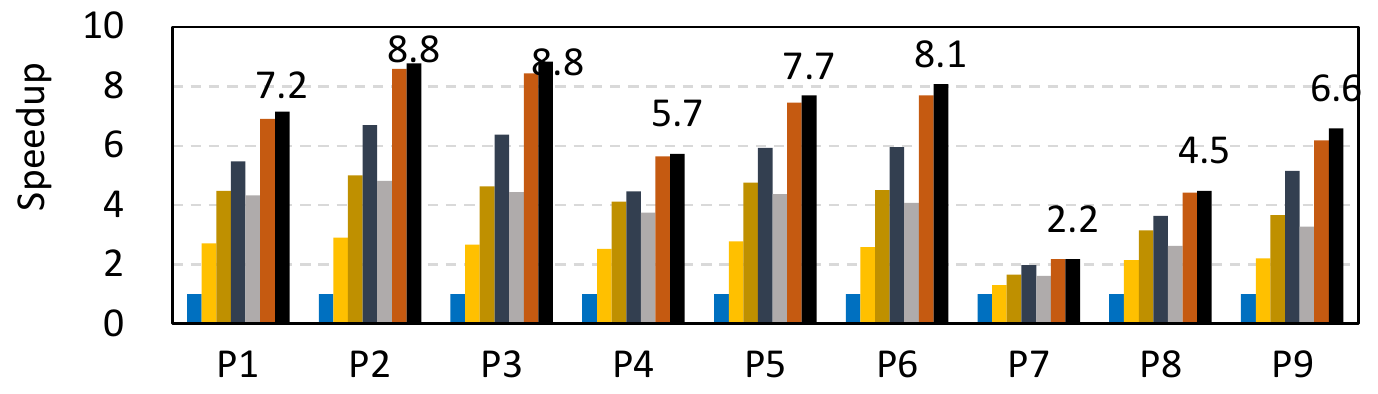}
 \label{fig:2630}
 }
 \vspace{-0.5\baselineskip}
 \subfloat[Dual Gold 6140 CPUs.] {
 \includegraphics[width=0.5\textwidth]{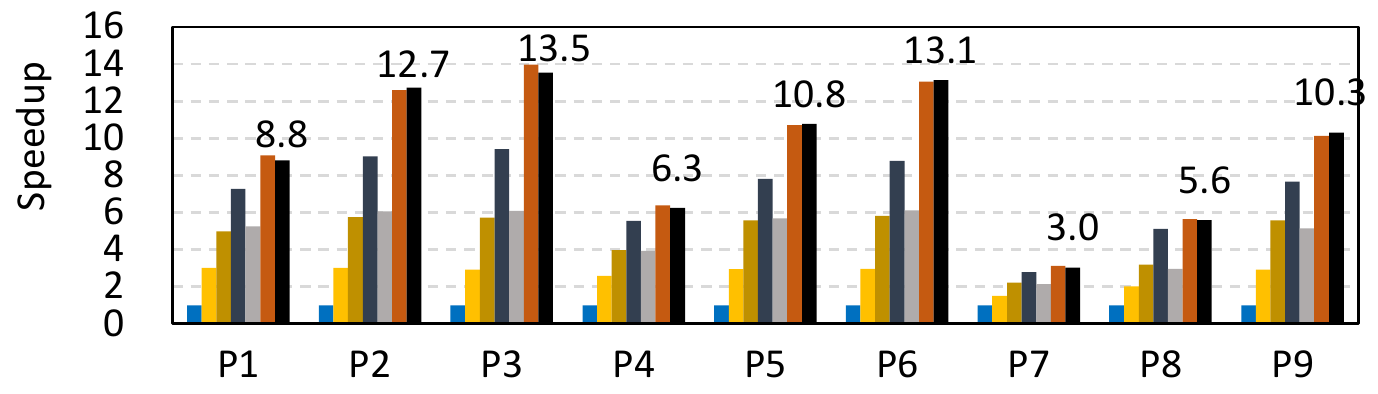}
 \label{fig:gold}
 }
 \subfloat[A single i7-9700K CPU.] {
 \includegraphics[width=0.5\textwidth]{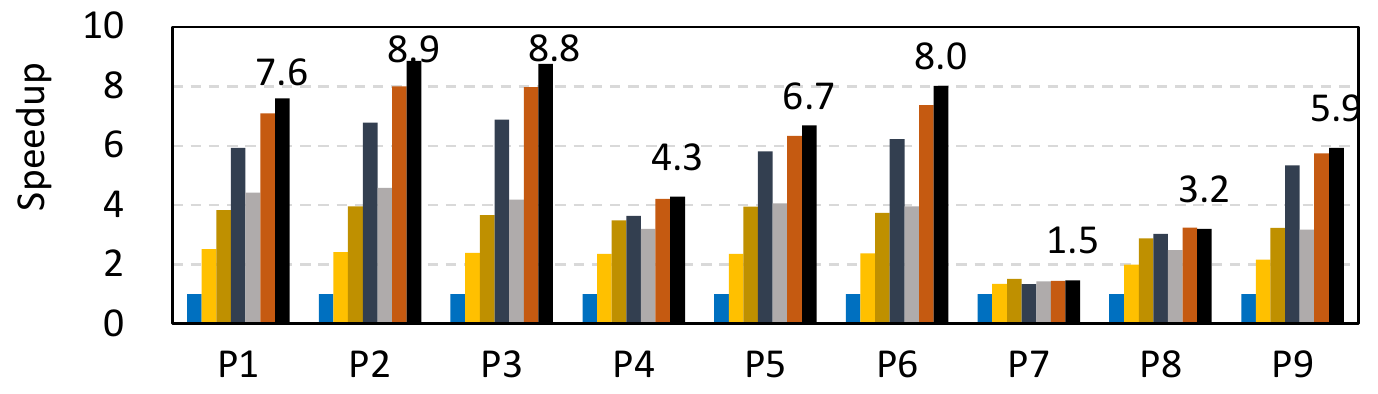}
 \label{fig:9700k}
 }
\caption{Speed up for OpenCL implementation (Intel CPUs) at different levels of optimizations.}
\label{fig:ocl-performance}
\end{figure*}
\begin{figure}[t]
  \begin{center}
    \includegraphics[clip,width=0.495\textwidth]{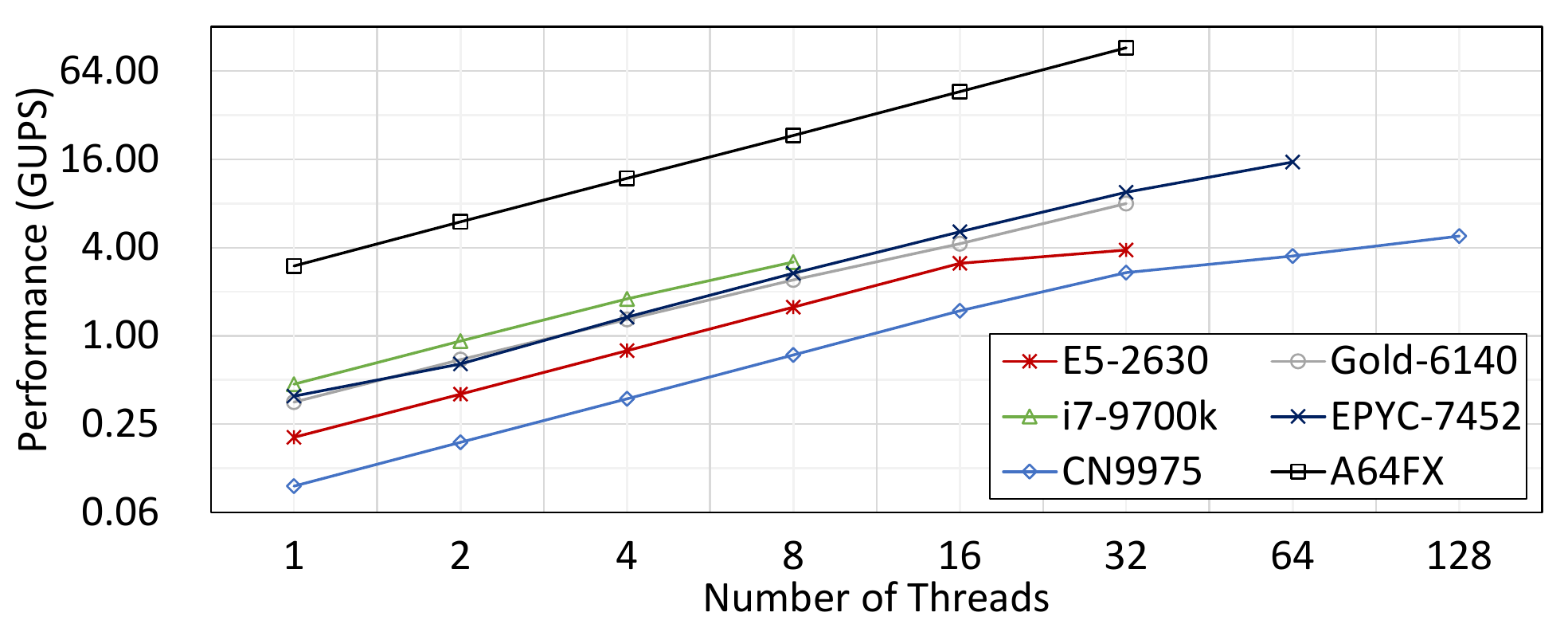}
    \caption{Scaling of the proposed back-projection with the number of threads for problem P5 ($512\times512\times512\Rightarrow512^3$).}
    \label{fig:scaling-baseline}
  \end{center}
\end{figure}

\subsection{Impact of Reducing Memory Accesses}
This section discusses the performance impact of reducing memory accesses by batched processing. 
There are four kernels that reduce the memory accesses, namely \emph{subline\_mp}, \emph{share\_lm\_cl}, \emph{symmetry\_lm\_cl}, and \emph{symmetry\_pf\_cl} (the top performing kernel that includes all optimizations).
In Figure~\ref{fig:batch-number}, we show the performance of \emph{symmetry\_pf\_cl} with different configurations values for the batch numbers going: $1, 2, 4, \dots, 32$. 
The total number of memory accesses to projections (in float32 words) is $N_{memProj}{\approx}4*nx*ny*nz*np$ (four accesses per voxel update by bilinear interpolation). The total number of memory accesses to the volume data is $N_{memVol}=nx*ny*nz*np/{nb}$, as explained in Section~\ref{sec:reduc-memory-access}. Therefore, the total number of memory accesses are
\begin{equation*}\begin{split}\label{equ:total_mem} 
N_{mem} = N_{memProj} + N_{memVol} \approx (4+\frac{1}{nb})*np*nx*ny*nz  
\end{split}\end{equation*}
For a given image reconstruction problem, the variables np, nx, ny, and nz are fixed. Consequently, the performance is inversely proportional to $N_{mem}$.
When observing the performance in Figure~\ref{fig:batch-number}, the performance behavior demonstrates this inverse proportional relation to equation $N_{mem}$.

In addition, we observe a performance gain when using a larger batch number for most of the image reconstruction cases. Since the performance of back-projection is bounded by the memory bandwidth, the proposed algorithm benefits from the reduced memory access when batching the projections.
To simplify the parameters configuration, in the following experiments we report results for a fixed batch number of \emph{32}.

\subsection{Performance \& Scalability on CPUs}
This section reports the performance and scalability of the proposed algorithms.
Figure~\ref{fig:batch-speedup} and Figure~\ref{fig:ocl-performance} show the performance of several back-projection kernels on different CPUs and image reconstruction problems as listed in Table~\ref{tbl:problems}.
The levels of optimization of all evaluated kernels can be found in Table~\ref{tbl:kernel-names}.
We show speedup over the baseline, which is the multi-threaded and widely used back-projection implementation in the RTK library.
The OpenMP-optimized \emph{subline\_mp} and OpenCL-based \emph{symmetry\_pf\_cl} achieve the highest performance in most of the cases, due to the collective use of optimizing techniques. 
The speedup of \emph{subline\_mp} ranges from \emph{0.3} to \emph{5.6} and \emph{symmetry\_pf\_cl} ranges from \emph{1.5} to \emph{13.1}. Upon investigation, it became clear that this performance differential between OpenMP and OpenCL is due to the higher quality of vectorized code by the OpenCL compiler.
Taking the \emph{subline} kernel as illustrated in Section~\ref{sec:vectorization} for instance, the assembly snippet can be found in Listing~\ref{alg:load_mix}. The quality of those vectorized codes (AVX2) is competitive to the hand-optimized ones that we tested.
To sum up, the portable performance improvement of the proposed algorithms is due to better vectorization, improved data locality, and reduced memory traffic.

Figure~\ref{fig:scaling-baseline} shows the scalability of the proposed back-projection algorithm (\emph{subline\_mp}) on several CPUs (listed in Table~\ref{tbl:evaluation-env}). The number of threads is configured as the power of two and ranges between \emph{1} to \emph{128}. The performance on most CPUs scales linearly to the number of cores up to some point where we observe saturation of the memory bus.

\begin{figure}[t]
  \begin{center}
    \includegraphics[clip,width=0.495\textwidth]{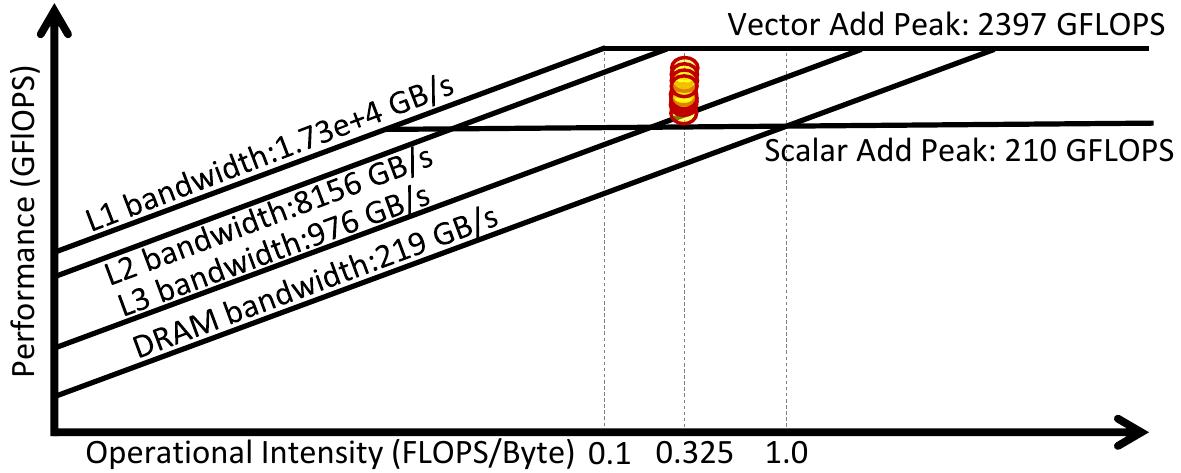}
    \caption{Roofline analysis of \emph{symmetry\_pf\_cl}, provided by Intel advisor profiler~\cite{marques2017performance,intel:advisor} on dual Gold-6140 CPUs. The problem size P1, P2, $\dots$, P9 are shown as yellow dots.}
    \label{fig:roofline}
  \end{center}
\end{figure}
\input{listings/code-load-mix}

\subsection{{Discussion}}
We show the speedup of several kernels in both Figure~\ref{fig:batch-speedup} and Figure~\ref{fig:ocl-performance} by comparing it with the baseline kernel. We observe the following:
{
    \setlength{\leftmargini}{15 pt}
\begin{enumerate*}[label=(\Roman*)]
\item OpenCL-based kernels outperform the OpenMP-based kernels due to the difference in the quality of the vectorized codes. The highest speed up among OpenMP-optimized kernels is almost $5.6\times$, in comparison to $13.5\times$ for OpenCL-optimized kernels.
\item Prefetching is effective in improving the performance in several kernels: \emph{symmetry\_pf\_cl} (kernel using prefetching) performs better than \emph{symmetry\_cl} (kernel not using prefetching) across different CPUs and image reconstruction problem sizes. The performance gain is noticeable specially in Gold-6140 CPU, likely due to that CPU's large L2/L3 cache (as listed in Table~\ref{tbl:evaluation-env}).
\item The bilinear interpolation scheme (in Section~\ref{sec:sub-line}) is very effective in improving the performance of back-projection kernels. It is clear that both \emph{share\_lm\_cl} and \emph{symmetry\_lm\_cl} (kernels using the subline bilinear scheme) perform better than \emph{share\_cl} and \emph{symmetry\_cl} (kernels not using the subline bilinear scheme), respectively.  
\item The optimization techniques such as shared reuse of hoisted variables and exploiting geometry symmetry (illustrated in Section~\ref{sec:opt-bp}) contribute to improving the performance of back-projection kernels. The kernels \emph{share\_cl}, \emph{share\_lm\_cl}, and \emph{symmetry\_cl} outperform the kernel \emph{transposal\_cl} that only transposes the projection and volume.
\item As shown in Figure~\ref{fig:roofline}, we display the Roofline model for the kernel of the top-performing kernel \emph{symmetry\_pf\_cl} on dual Gold-6140 CPUs. The optimizations we introduce push the effective achievable bandwidth to be between the bandwidths of L3 and L2 caches.
\end{enumerate*}
}

\begin{figure}[t]
  \begin{center}
    \includegraphics[clip,width=0.495\textwidth]{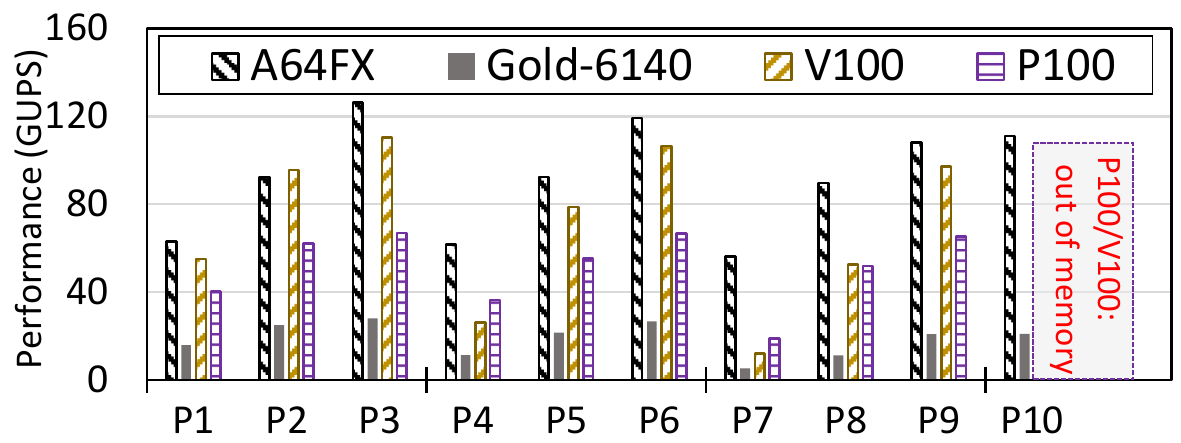}
    \caption{Performance comparison with GPUs at single-precision (end-to-end image reconstruction). 
    Problem P10 can not be solved on P100/V100 GPUs since it exceeds the capacity of device memory. For GPUs, we use the CUDA implementation in the RTK library.}
    \label{fig:performance:cpu:gpu}
  \end{center}
\end{figure}
\subsubsection{\bf{Comparison with GPUs}}
We demonstrate the competitive performance of our back-projection implementation on the A64FX ARM CPU by comparing it to the high-end Nvidia P100~\cite{nvidia2016p100} and V100~\cite{nvidia2017corporation} GPUs (see Figure~\ref{fig:performance:cpu:gpu}).
The RTK library implementation for GPUs
is used for our comparison. 
RTK is one of the most widely used open-source library implementations on GPUs, to the author's knowledge. Our comparison includes the overhead of moving the projections from host memory to device memory and excludes moving the volume back to the CPU. Note that production-level GPU implementations of back-projection (including RTK) do not overlap the computation with moving the projections to the device. That is primarily because overlapping schemes would require complex alterations in the image reconstruction algorithm in addition to being ineffective when scaling the size of projections (i.e. the algorithm cannot function on partial projections).

To conduct a fair comparison, we perform all computations in single precision (without using mixed precision on GPUs or CPUs).
Using the A64FX CPU, we can achieve better performance than Tesla P100/V100 GPUs on a wide range of image reconstruction problems. For fairness, we compare our work to RTK's CUDA implementation since RTK is one of the most used open libraries in both research and industry, though RTK's performance can be further optimized using the algorithms in~\cite{lu2016cache}. Note that the implementation in~\cite{lu2016cache} is not publicly available.
The RTK implementation uses a double buffer method to store volume data (read data from a volume buffer and write the updated value to another buffer). Hence, the maximum capacity of the generated volume must be less than 8 GB. That is because the device memory is limited to 16 GB in P100/V100 GPUs, and a fraction of the device memory is also required to store the projections. It is important to mention that the A64FX can solve the bigger problems such as P10 ($1024^2\times512\Rightarrow1300^3$) in Figure~\ref{fig:performance:cpu:gpu}, however, the P100/V100 GPUs are restricted by their memory capacities. 

To sum up, with the A64FX CPU we reach performance that rivals the top-performing GPUs when accounting for the unavoidable data movement overhead.

\section{Related work}\label{sec:related work}
Tomographic image reconstruction has been heavily researched in the past years.
Target-specific hardware is often adopted to speed up the computation of back-projection. Swindell et al.~\cite{swindell1983computed} proposed a linear accelerator to optimize the back-projection for the first generation CT devices three decades ago. Wu et al.~\cite{wu1991asic} used Application-Specific Integrated Circuits (ASIC) to speed up the back-projection algorithm.
The authors in~\cite{coric2002parallel,xue2006acceleration,subramanian2009c,henry2012fpga} also employed an FPGA to tune the computation of the FDK algorithm.
Several authors also use high-level synthesis methods~\cite{coric2002parallel,xue2006acceleration,subramanian2009c,henry2012fpga}, e.g. OpenCL, to generate the fast back-projection kernels on FPGAs rather than using HDL language.
Treibig et al.~\cite{treibig2013pushing} employed the SIMD instruction set extensions (coded in assembly) to speed up the back-projection computation and achieved outstanding performance on CPUs, e.g. using four Xeon E7-4870 CPU to generate a volume of size $1024^3$ and achieve the performance of $12$ GUPS. Notably, such performance is less than a quarter of the peak performance on Intel CPUs. Furthermore, Treibig et al. took advantage of the gather load intrinsic to alleviate the pressure of memory traffic on accessing projections (e.g. access along line $\bar{a}$ in Figure~\ref{fig:geometry-motivation}). However, the performance of the optimized kernel was still bounded by the memory accesses to update the volume data. Unlike their method, we can fully vectorize the back-projection and reduce the memory access for projection and volume data.
In~\cite{hofmann2014performance}, Johannes et al. optimized the \emph{RabbitCT} benchmark on the Intel Xeon Phi accelerator. Using a fixed-point DSP (Digital Signal Processor) platform, Liang et al. presented an optimized implementation of FDK and achieved state-of-art balance in cost and power consumption~\cite{liang2010optimized}.
On several hardware accelerator platforms such as CPUs, GPGPUs, and Intel Xeon Phi, Serrano et al.~\cite{serrano2014high} proposed a parallelized FDK implementation.

Xiao et al.~\cite{xiao:mbir16} proposed a super-voxel technique to improve the data locality of the MBIR algorithm. However, super-voxels were restricted for parallel-beam based CT.
Mert et al.~\cite{hidayetoglu2020petascale} optimized an iterative image reconstruction algorithm and achieved petaflops performance on Summit supercomputer for parallel-beam CT systems, yet their solution is difficult to be applied to the current generation of CT devices (namely Cone-beam CT). In this work, we target Cone-beam datasets ($7^th$ generation CT) for which voxel-based back-projection is required. 
Lu et al.~\cite{lu2016cache} proposed a highly optimized back-projection algorithm for out-of-core computations. Their implementation is more specific for CUDA, e.g. using texture memory. Furthermore, the projection computation in our algorithms is fully different from their approach, e.g. we use the 3$\times$4 projection matrix shown in Listing~\ref{alg:bp-baseline}.

RTK library (the baseline of this work) is a widely used library for image reconstruction and provides implementations of back-projection kernels for both CPU and GPU architectures. Due to the complexity of overlapping schemes in back-projection, the unavoidable overhead of data movement between host and device may degrade the performance. Furthermore, the GPU memory is limited and thus, volume decomposition techniques are required to go out-of-core~\cite{lu2016cache,biguri2019arbitrarily}.

\section{Conclusion} \label{sec:conclusion}

We revisit the role of CPUs in image reconstruction, motivated by the considerations of cost, power, and space requirements. To use CPUs for compute-intensive medical image processing applications, this paper proposes performance portable back-projection algorithms and demonstrates the performance/scaling
on a wide range of multicore processors. 
Our implementations benefit from the agnostic vectorization, improved memory access pattern, and reduced arithmetic computations. We also propose a bilinear interpolation algorithm to cache the sub-line values of projections for reducing the memory access and the arithmetic computations. 
Our results show that the proposed back-projection can achieve, on average, 5.2$\times$ speedup over the multi-threaded implementation of the most widely used library, on a wide variety of CPUs. We demonstrate the capability of using an ARM CPU (A64FX) to outperform high-end GPUs in the domain of CT, which is traditionally driven by GPUs.
\section*{Acknowledgment}
This work was supported by JSPS KAKENHI Grant Number JP21K17750.
This work was partially supported by JST-CREST under Grant Number JPMJCR19F5; JST, PRESTO Grant Number JPMJPR20MA, Japan.
We would like to thank Endo Lab at Tokyo Institute of Technology for providing computing resources.
The author wishes to acknowledge useful discussions with Dr. Jintao Meng at Chinese Academy of Science (CAS).
\bibliographystyle{ACM-Reference-Format}
\bibliography{main}

\end{document}

%% file: listings/alg-bp-baseline.tex
\lstset{
 	language = C++, breaklines = true, breakindent = 10pt, basicstyle = \rmfamily\footnotesize, commentstyle = {\itshape \color[cmyk]{1,0.4,1,0}}, classoffset = 0, keywordstyle = {\bfseries \color[cmyk]{0,1,0,0}}, stringstyle = {\ttfamily \color[rgb]{0,0,1}}, frame = trbl, framesep=5pt, numbers = left, stepnumber = 1, xrightmargin=7pt, xleftmargin=8pt, numberstyle = \tiny, tabsize = 3, captionpos = t, directivestyle={\color{black}},  emph={int,char,double,float,unsigned}, emphstyle={\color{blue}}
}
\lstset{escapeinside={<@}{@>}}

\begin{figure}[t]
\centering
\begin{minipage}[c]{0.5\textwidth}
\begin{lstlisting}[caption = {Pseudocode of the multi-threaded back-projection implementation adopted by RTK and RabbitCT. The {\bf{np}} is the number of projections. The sizes of the 2D projections, 3D volume data, and projection matrices are {\bf{nh$\times$nw}}, {\bf{nz$\times$nx$\times$ny}}, and {\bf{3$\times$4}}, respectively.}, label = alg:bp-baseline]
void <@\bf{BackProjection}@>(float img[np][nh][nw], float mat[np][3][4], float volume[nz][ny][nx])
{
   for (int s = 0; s < np; s++){
	 <@\bfseries\textcolor[cmyk]{0,1,0,0}{parallel\textunderscore for}@> (int k = 0; k < nz; k++){
	   for (int j = 0; j < ny; j++){
	    for (int i = 0; i < nx; i++){
	        float vec[4] = {i,j,k,1.f};//coordinate 
			  float z = <@\bf{dot4}@>(mat[s][2], vec); //dot
			  float f = 1.f/z;
			  float x = <@\bf{dot4}@>(mat[s][0], vec)*f; //dot
			  float y = <@\bf{dot4}@>(mat[s][1], vec)*f; //dot  
			  float val = <@\bf{Bilinear\_Interpolate}@>(img[s], x, y);
			  float w = f*f;  //compute weight
			  volume[k][j][i] += val*w;//update
   } } } } // s, k, j, i
}
\end{lstlisting}
\end{minipage}
\end{figure}

%% file: listings/alg-dot4.tex
\lstset{
 	language = C++, breaklines = true, breakindent = 10pt, basicstyle = \rmfamily\footnotesize, commentstyle = {\itshape \color[cmyk]{1,0.4,1,0}}, classoffset = 0, keywordstyle = {\bfseries \color[cmyk]{0,1,0,0}}, stringstyle = {\ttfamily \color[rgb]{0,0,1}}, frame = trbl, framesep=5pt, numbers = left, stepnumber = 1, xrightmargin=7pt, xleftmargin=8pt, numberstyle = \tiny, tabsize = 2, captionpos = t, directivestyle={\color{black}},  emph={int,char,double,float,unsigned, Type}, emphstyle={\color{blue}},
}
\lstset{escapeinside={<@}{@>}}

\begin{figure}[t]
\centering
\begin{minipage}[c]{0.5\textwidth}
\begin{lstlisting}[caption = {The Bilinear\_Interpolate is a bilinear interpolation function. The \emph{dot4} is an inner product function and the \emph{mix} is a built-in common function (e.g. in OpenCL) that returns a linear blend of two variables, where \emph{Type} can be scalar or vector data types. }, label = alg:subpixel]
float <@\bf{Bilinear\_Interpolate}@>(float img[nh][nw], float x, float y){
  int nx = (int)x; //convert to integer
  int ny = (int)y; //convert to integer
	//horizontal interpolation using mix
	float s0 = <@\bf{mix}@>(img[ny  ][nx], img[ny  ][nx+1], x-nx);
	float s1 = <@\bf{mix}@>(img[ny+1][nx], img[ny+1][nx+1], x-nx);
	//vertical interpolation using mix
	return <@\bf{mix}@>(s0, s1, y-ny);
}
float <@\bf{dot4}@>(float v0[4], float v1[4]){//inner product   
    return v0[0]*v1[0]+v0[1]*v1[1]+v0[2]*v1[2]+<@\bf{v0[3]}@>;
}
Type <@\bf{mix}@>(Type x, Type y, Type a){//linear interpolation   
    return x*((Type)1.f-a) + y*a;
}
\end{lstlisting}
\end{minipage}
\end{figure}


%% file: algorithm/alg-scratchpad-subline.tex
\begin{algorithm}[t]
      \small
\caption{Proposed back-projection. \emph{nb} is the batch number. The transposed projection and volume data are presented as \emph{img} and \emph{volume}, respectively.}
\label{alg:scratchpad-subline}   
\LinesNumbered
\Input{img[nb][nw][nh], mat[nb][3][4], volume[nx][ny][nz], vecIJ\Comment{the pairs of i and j are stored in vecIJ}}
\Output{volume[nx][ny][nz]}

{\color{blue}\#pragma omp parallel for} \Comment{launch multiple threads}

\ForEach {\{i, j\} $\in$ vecIJ}{

    vec $\gets$ \{i, j, 0, 1.f\} \Comment{k is independent of X}

    \For{$s=0$ to $nb-1$}{
        F[s]{$\gets$}1.f/dot4(mat[s][2], vec)\Comment{an array of f}
        
        W[s]{$\gets$}F[s]*F[s]               \Comment{an array of w}        
        
        X[s]{$\gets$}dot4(mat[s][0], vec)*F[s]\Comment{an array of x}
    }
    
    \For{$s=0$ to $nb-1$}{
        nx{$\gets$}(int)X[s] \Comment{type convension}
    
        {\color{blue}\#pragma omp simd}
        
        \For{m=0 to nh-1}{
        
            sMem[s][m]$\gets$mix(img[s][nh][m], img[s][nh+1][m], X[s]-nX)\Comment{\bf see Figure~\ref{fig:subline-algorithm}}
        }
    }

    \For{$k=0$ to $\frac{nz}{2}-1$}{
        
        sum$\gets$0 \Comment{register for partial results}
        
        \_sum$\gets$0 \Comment{register for partial results}
        
        vec[2]$\gets$k;\Comment{update index k}
        
        {\color{blue}\#pragma omp simd}
        
        \For{$s=0$ to $nb-1$}{
            pMem$\gets$sMem[s]
            
            y$\gets$dot4(mat[s][1], vec)*F[s]\Comment{projection comp.}
            
            ny$\gets$((int)y\Comment{type conversion}
        
           sum$\pluseq$mix(pMem[ny], pMem[ny+1], y-ny)*W[s]
           
           y$\gets$nh - 1 - y\Comment{geometry symmetry}
           
           ny$\gets$(int)y\Comment{type conversion}
           
           \_sum $\pluseq$ mix(pMem[ny], pMem[ny+1], y-ny)*W[s]
           
        }
        
        volume[i][j][k] $\pluseq$ sum\Comment{update volume data}
        
    	volume[i][j][nz-1-k] $\pluseq$ \_sum\Comment{update volume data}
    }

}
\end{algorithm}

%% file: figures/alg-subline.tex
\lstset{
 	 breaklines = true, breakindent = 10pt, basicstyle = \ttfamily\small, commentstyle = {\itshape \color[cmyk]{1,0.4,1,0}}, classoffset = 0, keywordstyle = {\bfseries \color[cmyk]{0,1,0,0}}, stringstyle = {\ttfamily \color[rgb]{0,0,1}}, frame = trbl, framesep=5pt, numbers = left, stepnumber = 1, xrightmargin=7pt, xleftmargin=8pt, numberstyle = \ttfamily\scriptsize, tabsize = 3, captionpos = t, directivestyle={\color{black}},  
}
\lstset{escapeinside={<@}{@>}}






\begin{figure*}[t]
\centering
\begin{minipage}[c]{0.485\textwidth}
\subfloat[\emph{ptr0} and \emph{ptr1} are pointers to two neighbouring vertical rows. \emph{sMem} is a small-sized memory buffer for caching the sub-line values, where 0$\leq$dx$<1$.] {
    \includegraphics[width=1\textwidth]{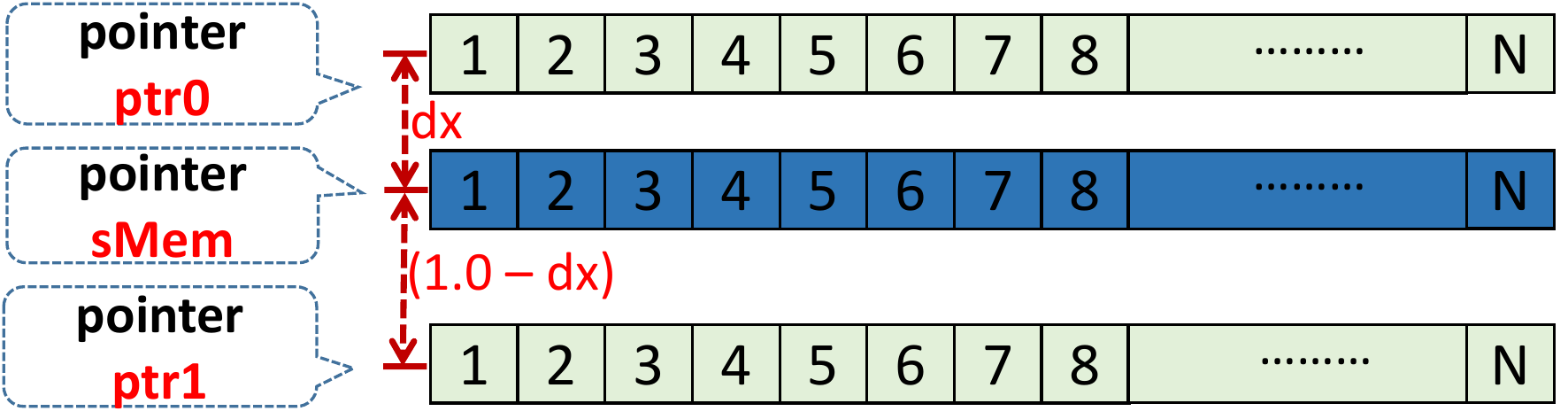}
    \label{fig:subline}
 }
\end{minipage}
\hfill\hspace{0.005\textwidth}
\begin{minipage}[c]{0.485\textwidth}
\subfloat[The bilinear interpolation in the horizontal direction. Each output value rely on two neighboring values at the sub-line buffer.] {
    \includegraphics[width=1\textwidth]{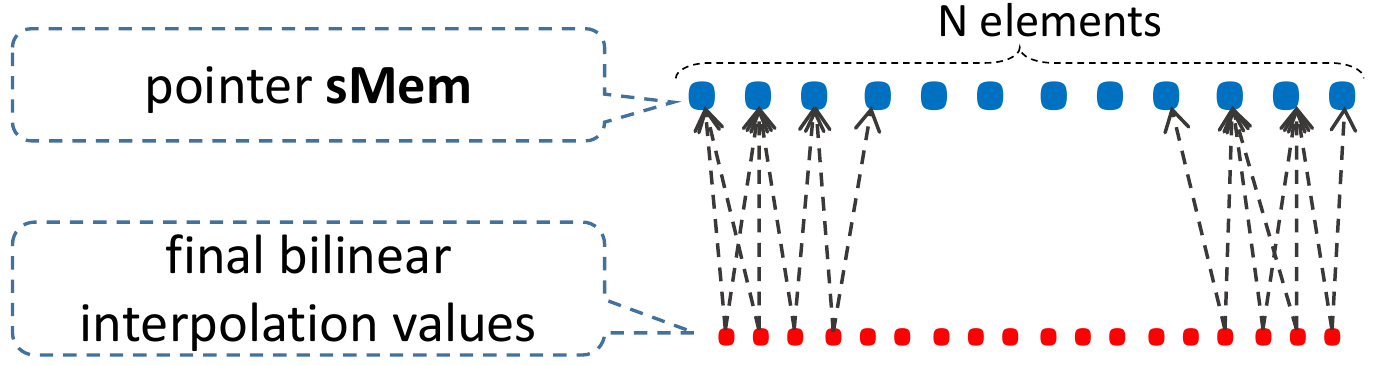}
    \label{fig:use-subline}
 }
\end{minipage}
\caption{Novel interpolation scheme using sub-line blending. (a) and (b) show vertical and horizontal interpolations.
}
\label{fig:subline-algorithm}
\end{figure*}

\begin{figure*}[t]
\centering
\begin{minipage}[c]{0.47\textwidth}
\subfloat[Vectorize sub-line alg. by C (OpenMP).] {
\begin{lstlisting}[firstline=0]
#pragma omp simd
for (int i=0; i<N; i++)
{
    //blend two lines of pixels
    sMem[i] = <@\bf{mix}@>(ptr0[i], ptr1[i], dx)
}
\end{lstlisting}
    \label{fig:subline-by-c}
}
\end{minipage}
 \hfill\hspace{0.02\textwidth}
\begin{minipage}[c]{0.47\textwidth}
\subfloat[
Vectorize sub-line alg. by OpenCL.
] {
\begin{lstlisting}[firstline=0, lastline=6]
for (int i=0; i<N/8; i ++){
   float8 v0 = <@\bf{vload8}@>(i, ptr0);     //load float8
   float8 v1 = <@\bf{vload8}@>(i, ptr1);     //load float8
   float8 v2 = <@\bf{mix}@>(v0, v1, dx);    //blend float8
   <@\bf{vstore8}@>(v2, i, sMem);             //store float8
}
\end{lstlisting}
    \label{fig:subline-by-opencl}
}
\end{minipage}
\caption{Implementations of proposed bilinear interpolation scheme. \emph{mix} is a linear interpolation function in Listing~\ref{alg:subpixel}.
}
\label{fig:subline-algorithm-code}
\end{figure*}

%% file: listings/alg-ocl-bp-symmetry-lm.tex
\lstset{columns=fullflexible,
 	 breaklines = true, breakindent = 7pt, basicstyle = \rmfamily\footnotesize, commentstyle = {\itshape \color[cmyk]{1,0.4,1,0}}, classoffset = 0, keywordstyle = {\bfseries \color[cmyk]{0,1,0,0}}, stringstyle = {\ttfamily \color[rgb]{0,0,1}}, frame = trbl, framesep=5pt, numbers = left, stepnumber = 1, xrightmargin=7pt, xleftmargin=8pt, numberstyle = \tiny, tabsize = 3, captionpos = t, directivestyle={\color{black}},  emph={__global, __local, __kernel, int,char,double,float,unsigned, int2, int3, float2, float4}, emphstyle={\color{blue}},
}
\lstset{escapeinside={<@}{@>}}

\begin{figure}[t]
\centering
\begin{minipage}[c]{0.5\textwidth}
\begin{lstlisting}[caption = {A performance-portable OpenCL kernel for back-projection implementing our optimization techniques, e.g. transposing projection and volume data, reordering loops, reusing variables after hoisting, and exploiting geometric symmetry. Constant memory is used to store the projection matrix (see \emph{mat}), global memory for the projections (see \emph{img}) and volume data (see \emph{volume}), where ROWS$\equiv$3, imgDim=(np, nw, nh), volDim=(nx, ny, nz), LM\_SIZE=imgDim.x=nh.}, label = alg:ocl-bp-symmetry-lm]
__kernel void bp_optimized( __constant float4* mat, __global float* img, 
    int3 imgDim, __global float* volume, int3 volDim, __global int2* vecIJ)
{
	int k = <@\bf{get\_global\_id}@>(0);          //global index x
	__local int2 ij; //share index i and j
	__local float F[nb], X[nb], W[nb], sMem[LM_SIZE];//local mem
	if (k == 0) ij = vecIJ[<@\bf{get\_global\_id}@>(1)]; //i and j <@$\leftarrow$@> global index y
	barrier(CLK_LOCAL_MEM_FENCE); //barrier for local memory

	float4 ijkw = (float4)(ij.x, ij.y, k, 1.f); //vector (i,j,k)
	if (k < nb) {
		float z = 1.f / dot((mat + ROWS * k)[2], ijkw);//compute z
		X[k] = dot((mat + ROWS * k)[0], ijkw)*z; //compute x
		F[k] = z;                                //cache z
		W[k] = z * z;                           //compute weight
	}
	barrier(CLK_LOCAL_MEM_FENCE); //barrier for local memory

	float2 sum = (float2)(0.f, 0.f);
	for (int s = 0; s < nb; s++, img += imgDim.x∗imgDim.y;, mat += ROWS){
		int nx = convert_int(X[s]);
		float dx = X[s] - convert_float(nx);
		__global float* ptr0 = img + nx * imgDim.x; <@\bf\color{red}{  //see Fig.~\ref{fig:subline}}@>
		__global float* ptr1 = ptr0 + imgDim.x;       <@\bf\color{red}{  //see Fig.~\ref{fig:subline}}@>
		for (int m = k; m < LM_SIZE; m += <@\bf{get\_local\_size}@>(0)) 
			<@\hllightgray{sMem[m] = mix(ptr0[m], ptr1[m], dx);}@>  <@\bf\color{red}{  //see Fig.~\ref{fig:subline-algorithm-code}}@>
		barrier(CLK_LOCAL_MEM_FENCE); //barrier for local memory
		
		float y = dot(mat[1], ijkw)*F[s];
		float _y = width - 1 - y; //y and _y are symmetric at the vertical line of FPD  
		int2 ny = convert_int2((float2)(y, _y));//float <@$\rightarrow$@> int
		float2 dy = (float2)(y, _y) - convert_float2(ny);
		<@\bf\color{red}{//linear interpolation and update sum as in Fig.~\ref{fig:use-subline}}@>
		<@\hllightgray{sum += mix((float2)(sMem[ny.x], sMem[ny.y]), (float2)(sMem[ny.x+1], sMem[ny.y+1]), dy)*W[s];}@>
		barrier(CLK_LOCAL_MEM_FENCE); //barrier for local memory
	}
	int offset = ij.y*volDim.z*volDim.x + ij.x * volDim.z;
	volume[offset + k] += sum.x; //update volume
	volume[offset + volDim.z - 1 - k] += _sum.y; //update volume
}
\end{lstlisting}
\end{minipage}
\end{figure}


%% file: tables/tbl-ComputeDevice.tex
 \begin{table*}[t]  
      \caption{Evaluation environment.}      
      \footnotesize
      \centering
      \resizebox{\linewidth}{!}{
      \begin{tabular}{ ?c|c|c|c|c|c|c|c? }
      \tbhline
      \multirow{2}{*}{\bf{CPU Name}}  & \bf{Intel Xeon}    & \bf{Intel Xeon}  &\bf{Intel Xeon}  & \bf{Intel Core}   &  \bf{AMD}      &  \bf{ThunderX2 (ARM)} &  \bf{Fujitsu (ARM)}\\
       \bf{}                     & \bf{E5-2630 v4}    & \bf{E5-2650 v3}  &\bf{Gold 6140}   & \bf{i7-9700K}     &  \bf{EPYC-7452}&  \bf{CN9975} &  \bf{A64FX} \\
       \hline
          Cores                  &      10            & 10               &18               & 8                 & 32           & 28          &  48        \\
          Threads/core           &      2             & 2                & 1               & 1                 & 1            & 4           &  1         \\
          Frequency              &   {2.2GHz}         & {2.3GHz}         &{2.3GHz}         & {3.6GHz}          & 2.35GHz      & 1.8GHz      &  2.2GHz    \\
          Sockets                &   {2}              & {2}              &{2}              & {1}               & 2            & 2           &  1         \\
          GFLOPS (SP)            &   704       & 736       &2,649     & 461        & 2,406 & 806  &  6,800     \\
          Max bandwidth          &   {137GB/s}        & {137GB/s}        &{250GB/s}        & {42GB/s}          & 205GB/s      & 159GB/s     & 1024GB/s   \\
          L1d/L1i cache          &   32K/32K          & 32K/32K          & 32K/32K         & 32K/32K           & 32K/32K      & 32K/32K     & 3M/3M      \\
          L2/L3 cache            &   256K/25M         & 256K/25M         & 1M/25M          & 256K/12M          & 512K/16M     & 256K/32M    & 32M/---    \\
          TDP                    &    85W             &  105W            & 140W            & 95W               & 155W         & 150W & $\sim$160W  \\
      \tbhline
      \end{tabular}}
      \label{tbl:evaluation-env} 
\end{table*}     


%% file: algorithm/alg-prefetch.tex
\begin{algorithm}[t]
\begin{minipage}{0.97\linewidth}
\small
\caption{Prefetching by dual local memory buffers. }
\label{alg:prefetch}
\LinesNumbered
\Input{$\dots\dots$\Comment{the input argument are same to Listing~\ref{alg:ocl-bp-symmetry-lm}}}
  
\Output{volume\Comment{the output volume}}

{\hllightgray{\_\_local sMem[2][LM\_SIZE]}\Comment{declear dual scratchpad memories}}

{{$\text{sMem[0]} \gets img[0]$}\Comment{prefetching by sub-line alg. as in line 26}}

{$sum \gets 0$\Comment{declear registers for accumulation}}

\For {$i = 0$ to $nb-1$}{
    \uIf{$i+1 < B$}{
        {{$\text{sMem[(i+1)\%2]} \gets img[i+1]$}\Comment{prefetching without barrier}}
    }
    {$\text{sum} \gets sMem[i\%2]$\Comment{accumulate result via subline}}
    
    {{barrier(CLK\_LOCAL\_MEM\_FENCE)}\Comment{barrier work-items}}
}
{$\text{volume} \gets sum$\Comment{update volume by batched processing}}
\end{minipage}
\end{algorithm}

%% file: tables/tbl-kernelNames.tex
 \begin{table}[t]  
      \caption{Back-projection kernel names and optimizations.}
      \footnotesize
      \centering
      \resizebox{\linewidth}{!}{
      \begin{tabular}{ ?c?l?c|c|c|c|c|c? }
      \tbhline
       {\bf{API}}              & {\bf{Name}}       &   \rotatebox{90}{{Transpose}} & \rotatebox{90}{{Share}} & \rotatebox{90}{{Symmetry}} & \rotatebox{90}{{Subline}} &  \rotatebox{90}{{LocalMem}} & \rotatebox{90}{{Prefetching}} \\
      \tbhline
                               & transpose\_mp     & $\checkmark$ &                 &              &              &              &                \\
      {OpenMP}                 & share\_mp         & $\checkmark$ &   $\checkmark$  &              &              &              &                \\
                               & symmetry\_mp      & $\checkmark$ &   $\checkmark$  & $\checkmark$ &              &              &                \\
                               & subline\_mp       & $\checkmark$ &   $\checkmark$  & $\checkmark$ & $\checkmark$ &              &                \\  
     \tbhline                  
                               & transpose\_cl     & $\checkmark$ &              &              &          &              &                \\
                               & share\_cl         & $\checkmark$ & $\checkmark$ &              &          &              &                \\
                               & share\_lm\_cl     & $\checkmark$ & $\checkmark$ &              &          & $\checkmark$ &                \\
     {OpenCL}                  & symmetry\_cl      & $\checkmark$ & $\checkmark$ & $\checkmark$ &          &              &                \\  
                               & symmetry\_lm\_cl  & $\checkmark$ & $\checkmark$ & $\checkmark$ &          & $\checkmark$ &                \\     
                               & symmetry\_pf\_cl  & $\checkmark$ & $\checkmark$ & $\checkmark$ &          & $\checkmark$ & $\checkmark$   \\   
      \tbhline
      \end{tabular}}  
      \label{tbl:kernel-names} 
\end{table}

%% file: tables/tbl-problems.tex
 \begin{table}[t]  
      \caption{Evaluated image reconstruction problem sizes.}        
      \small
      \centering
      \resizebox{\linewidth}{!}{
      \begin{tabular}{ ?c|l?c|l? }
      \tbhline
      \bf{label}     & Problem    & \bf{label}     & Problem     \\
      \tbhline
      P1 & $256^2\times512\Rightarrow256^3$   & P7 & $1024^2\times512\Rightarrow256^3$    \\
      P2 & $256^2\times512\Rightarrow512^3$   & P8 & $1024^2\times512\Rightarrow512^3$    \\
      P3 & $256^2\times512\Rightarrow1024^3$  & P9 & $1024^2\times512\Rightarrow1024^3$   \\
      \hline
      P4 & $512^2\times512\Rightarrow256^3$   &     &                                      \\
      P5 & $512^2\times512\Rightarrow512^3$   & P10 & $1024^2\times512\Rightarrow1300^3$  \\
      P6 & $512^2\times512\Rightarrow1024^3$  &     &  ($1300^3$ volume is $\sim$8.2GB)   \\
      \tbhline
      \end{tabular}}
      \label{tbl:problems} 
\end{table}     

%% file: listings/code-load-mix.tex
\lstset{
 	 breaklines = true, breakindent = 10pt, basicstyle = \ttfamily\footnotesize, commentstyle = {\itshape \color[cmyk]{1,0.4,1,0}}, classoffset = 0, keywordstyle = {\bfseries \color[cmyk]{0,1,0,0}}, stringstyle = {\ttfamily \color[rgb]{0,0,1}}, frame = trbl, framesep=5pt, numbers = left, stepnumber = 1, xrightmargin=7pt, xleftmargin=8pt, numberstyle = \ttfamily\scriptsize, tabsize = 3, captionpos = t, directivestyle={\color{black}},  
}
\lstset{escapeinside={<@}{@>}}


\begin{figure}[t]
\centering
\begin{minipage}[c]{0.495\textwidth}
\begin{lstlisting}[backgroundcolor=\color{orange!30}, caption = {Assembly snippet from \emph{symmetry\_pf\_cl} OpenCL kernel with AVX2 vectorization.}, label = alg:load_mix]
0x177801a62b4   movsxd rdx, edx	                        
0x177801a62b7   vmovups ymm10, ymmword ptr [r8+rdx*4]
0x177801a62bd	vmovups ymm11, ymmword ptr [r14+rdx*4]
0x177801a62c3	vsubps ymm11, ymm11, ymm10	       
0x177801a62c8	vfmadd213ps ymm11, ymm1, ymm10	       
0x177801a62cd	mov rsi, rax	                        	
0x177801a62d0   shl rsi, 0xc	                 
0x177801a62d4	add rsi, r11
\end{lstlisting}
\end{minipage}
\end{figure}
